\documentclass[aps,prd,a4paper,onecolumn,amsmath,showpacs,superscriptaddress,nofootinbib,preprintnumbers,notitlepage]{revtex4-1}
\usepackage{amssymb,amsmath}
\usepackage{graphicx}
\usepackage{color}
\usepackage{dcolumn}
\usepackage{pgfplots}
\usepackage{booktabs}
\usepackage{multirow}
\usepackage{dcolumn}
\usepackage{amsmath}
\usepackage{mathtools}
\usepackage{amsfonts}
\usepackage{amssymb} 
\usepackage{epstopdf}
\usepackage{bm}
\usepackage{siunitx}
\usepackage{braket}
\usepackage{enumitem}
\usepackage{soul}
\usepackage{ulem}
\usepackage{color}
\usepackage{transparent}
\usepackage{pifont}
\usepackage[font={small}]{caption}
\usepackage{subfigure}
\usepackage{cancel}
\newcommand\rf[1]{(\ref{eq:#1})}
\newcommand\lab[1]{\label{eq:#1}}
\newcommand\nonu{\nonumber}
\newcommand\br{\begin{eqnarray}}
\newcommand\er{\end{eqnarray}}
\newcommand\be{\begin{equation}}
\newcommand\ee{\end{equation}}

\newcommand\lb{\lbrack}
\newcommand\rb{\rbrack}

\renewcommand\({\left(}
\renewcommand\){\right)}

\newcommand\bc{\begin{center}}
\newcommand\ec{\end{center}}

\newcommand\partder[2]{\frac{{\partial {#1}}}{{\partial {#2}}}}

\renewcommand\a{\alpha}

\renewcommand\d{\delta}

\newcommand\eps{\epsilon}
\newcommand\vareps{\varepsilon}

\newcommand\G{\Gamma}

\newcommand\h{\frac{1}{2}}
\renewcommand\k{\kappa}
\renewcommand\l{\lambda}

\newcommand\m{\mu}
\newcommand\n{\nu}
\newcommand\om{\omega}

\newcommand\p{\phi}
\newcommand\vp{\varphi}
\renewcommand\P{\Phi}
\newcommand\pa{\partial}

\renewcommand\th{\theta}

\newcommand\cL{{\mathcal L}}
\newcommand\cM{{\mathcal M}}

\newcommand\cP{{\mathcal P}}

\newcommand{\ct}[1]{\cite{#1}}
\newcommand{\bib}[1]{\bibitem{#1}}

\newcommand\PRL[3]{\textsl{Phys. Rev. Lett.} \textbf{#1} (#2) #3}

\newcommand\PRD[3]{\textsl{Phys. Rev.} \textbf{D#1} (#2) #3}

\newcommand\PLB[3]{\textsl{Phys. Lett.} \textbf{#1B} (#2) #3}
\newcommand\CQG[3]{\textsl{Class. Quantum Grav.} \textbf{#1} (#2) #3}

\newcommand\PR[3]{\textsl{Phys. Reports} \textbf{#1} (#2) #3}

\newcommand\IJMPD[3]{\textsl{Int. J. Mod. Phys.} \textbf{D#1} (#2) #3}

\newcommand\MPLA[3]{\textsl{Mod. Phys. Lett.} \textbf{A#1} (#2) #3}

\newcommand\vpdot{\stackrel{.}{\varphi}}
\newcommand\vpddot{\stackrel{..}{\varphi}}

\begin{document}

\title{Connecting Early Dark Energy to Late Dark Energy by the Diluting Matter Potential}

\author{Eduardo Guendelman}
\email{guendel@bgu.ac.il}
\affiliation{Department of Physics, Ben-Gurion University of the Negev, Beer-Sheva, Israel.\\}
\affiliation{Frankfurt Institute for Advanced Studies (FIAS),
Ruth-Moufang-Strasse 1, 60438 Frankfurt am Main, Germany.\\}
\affiliation{Bahamas Advanced Study Institute and Conferences, 
4A Ocean Heights, Hill View Circle, Stella Maris, Long Island, The Bahamas.
}

\author{Ram\'{o}n Herrera}
\email{ramon.herrera@pucv.cl}
\affiliation{Instituto de F\'{\i}sica, Pontificia Universidad Cat\'{o}lica de Valpara\'{\i}so, Avenida Brasil 2950, Casilla 4059, Valpara\'{\i}so, Chile.
}
\author{Pedro Labra\~{n}a}
\email{plabrana@ubiobio.cl}
\affiliation{Departamento de F\'{i}sica, Universidad del B\'{i}o-B\'{i}o, Casilla 5-C, Concepci\'on, Chile and\\
Centro de Ciencias Exactas, Universidad del B\'{i}o-B\'{i}o, Casilla 447, Chill\'an, Chile.}
\begin{abstract}
In this work we study a scale invariant gravity theory containing two scalar fields, dust particles and a measure defined from degrees of freedom independent of the metric. The integration of the degrees of freedom that define the measure spontaneously break the scale symmetry, leaving us in the Einstein frame with an effective potential that is dependent on the density of the particles. The potential contains three flat regions, one for inflation, another for early dark energy and the third for late dark energy. At a certain point, as the matter dilutes, tunneling from the early dark energy to the late dark energy can start efficiently. 
This mechanism naturally alleviated the observed Hubble tension by modifying the sound horizon prior to recombination while preserving late-time cosmology. Moreover, the model predictions are consistent with observations from the reduced CMB, BAO, and local measurement of $H_0$, providing a coherent and unified description of the universe. In this context, the Bayesian analysis of these datasets confirms the viability of our scenario, with the best-fit parameters indicating an early dark 
energy fraction of $f_{\rm NEDE}\approx 0.3$ at a redshift 
of $z^{\prime}=5000$. This preliminary estimate, obtained 
using the reduced CMB dataset, is expected to be tightened 
once the full CMB likelihood is considered.

\end{abstract}
\maketitle
\section{Introduction}
\label{intro}

In the standard cosmological framework for the early universe (see, for example, \ct{early-univ,primordial} and references therein), the universe begins with a period of rapid exponential expansion known as inflation.
Later, following the discovery of the accelerated expansion of the late universe \ct{accel-exp,accel-exp-2}, a similarly simple description emerged for the current cosmic evolution: the standard cosmological model for the late universe, commonly referred to as the $\Lambda$CDM model \ct{lambdaCDM}. This model includes a cosmological constant, dark matter, and ordinary visible (baryonic) matter.
According to this picture, the present universe is dominated by dark energy (DE), associated with the cosmological constant, which accounts for approximately 70$\%$ of the total energy density. This is followed by dark matter (DM), contributing about 25$\%$, while baryonic matter represents only about 5$\%$.

This simple $\Lambda$CDM is now being somewhat challenged by the discovery of several cosmological tensions, the most important being  the $H_0$  tension \ct{H0} followed by the $\sigma_8$ tension \ct{sigma8}. This suggests that the introduction of only a cosmological term to describe the DE and the addition of DM as dust, without any Dark Energy-Dark Matter interaction, for example, may be a too simple description of the post inflationary  Universe for the description of the Dark Energy and the Dark Matter. In addition to this DESI now present us with a tentative full history of the evolution of the DE, with a very interesting result that shows that the total equation of state (EoS) parameter  $w \approx -2 $ for $a \approx 0$, where $a$ is the expansion factor, see Ref.~\cite{reviewstrangeresultsbydesi}.

Now with the more recent results that show evidence of an $H_0$ tension, that  is a tension between the value of $H_0$ as derived from the supernova data and that derived from the CMB data, the early DE models have been suggested \cite{earlyde1, Niedermann:2020dwg}. 
In this context, the Hubble tension refers to the statistically significant discrepancy between the value of the Hubble constant $H_0$ inferred from early-universe observations, such as the Cosmic Microwave Background (CMB) measurements by \textit{Planck}, which suggest $H_0 \approx 67.4 \pm 0.5$ km s$^{-1}$ Mpc$^{-1}$ \cite{Planck:2018vyg}, and the higher values obtained from late-time, local measurements like those from the SH0ES project, reporting $H_0 \approx 73.30 \pm 1.0$ km s$^{-1}$ Mpc$^{-1}$ \cite{Riess:2021jrx}. Recent observations from the James Webb Space Telescope (JWST) have corroborated the higher local measurements, further intensifying the tension \cite{Riess:2023bfx}. This persistent discrepancy, now exceeding the $5\sigma$ level, suggests potential inadequacies in the standard $\Lambda$CDM model and has prompted the exploration of new physics, including early dark energy models \cite{Poulin:2023lkg} and modifications to the cosmic expansion history \cite{Khalife:2023qbu}. For a general review of the solutions of the $H_0$ problem, see \cite{Eleonora, WhitePaper}.

In this work, we investigate a potential mechanism to alleviate the Hubble tension within the framework of New Early Dark Energy (NEDE) models \cite{Niedermann:2019olb, Niedermann:2020dwg}. The NEDE is based on a first-order phase transition that occurs shortly before recombination
in a dark sector at zero temperature.
Theses models are motivated by the observation that Baryon Acoustic Oscillation (BAO) and Pantheon Supernova (SNe) data reveal a degeneracy between the Hubble constant $H_0$ and the sound horizon $r_s$, implying that $H_0 \propto \frac{1}{r_s}$, 
see \cite{earlyde1, Niedermann:2020dwg}.
Then, any cosmological framework attempting to accommodate a higher value of the Hubble constant $H_0$, while remaining consistent with CMB observations, must predict a reduced sound horizon $r_s$ at the drag epoch.
This constraint may suggests the presence of non-standard physics prior to recombination, as required to alter the early expansion history without conflicting with precision cosmological data.
In this context, the NEDE scenario offers a compelling mechanism by introducing a transient dark energy component that becomes dynamically relevant shortly before matter-radiation equality. This early injection of energy reduces the sound horizon and allows for a larger inferred value of $H_0$, thereby addressing the Hubble tension without invoking modifications to late-time cosmology.

Usually the NEDE scheme is realized by a quantum tunneling of a scalar field which is triggered at the right time (close to matter-radiation equality) by an additional sub-dominant trigger field, see Refs.~\cite{Niedermann:2019olb, Niedermann:2020dwg}. 
In our model, NEDE is also realized through the tunneling of a scalar field, however, the tunneling rate depends on the scale factor, which naturally triggers the phase transition without the need for additional fields.

A fundamental question remains unresolved, even before addressing the Hubble tension: how can we explain the existence of at least two epochs of exponential expansion—namely, the early inflationary phase and the current phase of late-time accelerated expansion—which occur at vastly different energy scales? Within our framework, this issue admits an elegant interpretation. Specifically, such behavior can be realized through a scalar field potential featuring two distinct and nearly flat regions. Furthermore, if we adopt the Early Dark Energy (EDE) hypothesis, a potential with three flat regions—corresponding to inflation, EDE, and late-time dark energy—can be constructed. Developing and exploring this scenario constitutes one of the central aims of this work.

The best known mechanism for generating a period of accelerated expansion 
is through the presence of some vacuum energy. In the context of a 
scalar field theory, vacuum energy density appears naturally when the scalar
field acquires an effective potential $U_{\rm eff}$ which has flat regions so 
that the scalar field can ``slowly roll'' \ct{slow-roll,slow-roll-param} and its 
kinetic energy can be neglected resulting in an energy-momentum tensor 
$T_{\m\n} \simeq - g_{\m\n} U_{\rm eff}$.

The possibility of continuously connecting an inflationary phase to a slowly 
accelerating universe through the evolution of a single scalar field -- the
{\em quintessential inflation scenario} -- has been first studied in 
Ref.~\ct{peebles-vilenkin}. Also, 
$F(R)$ models can yield 
both an early time inflationary epoch and a late time de Sitter phase with 
vastly different values of effective vacuum energies \ct{starobinsky-2}.
For a recent proposal of a quintessential inflation mechanism based on 
the k-essence  framework, see Ref.~\ct{saitou-nojiri}. For
another recent approach to quintessential inflation based on the 
``variable gravity'' model \ct{wetterich} and for extensive list of references 
to earlier work on the topic, 
see Ref.\ct{murzakulov-etal}. Other ideas based on the so called $\alpha$ attractors \ct{alphaatractors}, which uses non canonical kinetic terms have been studied. Also, a quintessential inflation based on a  Lorentzian slow-roll ansatz  which automatically gives two flat regions was studied in Ref.~\ct{Lorentzian}.

In previous  papers \ct{ourquintessence} we have studied 
a unified scenario where both an inflation 
and a slowly accelerated phase for the universe can appear naturally from the 
existence of two flat regions in the effective scalar field potential which
we derive systematically from a Lagrangian action principle. 
Namely, we started with a new kind of globally Weyl-scale invariant gravity-matter 
action within the first-order (Palatini) approach formulated in terms of two 
different non-Riemannian volume forms (integration measures) \ct{quintess}.
In this new theory there is a single scalar field with kinetic terms coupled to 
both non-Riemannian measures, and in addition to the scalar curvature 
term $R$ also an $R^2$ term is included (which is similarly allowed by global 
Weyl-scale invariance). Scale invariance is spontaneously broken upon solving part 
of the corresponding equations of motion due to the appearance of two 
arbitrary dimensionfull integration constants.

Let us briefly recall the origin of current approach. The main idea comes from 
Refs.~\ct{TMT-orig-1}-\ct{TMT-orig-3} (see also 
Refs.~\ct{TMT-recent-1-a}-\ct{TMT-recent-2}),
where some of us have proposed a new class of gravity-matter theories based on the 
idea that the action integral may contain a new metric-independent generally-covariant 
integration measure density, \textsl{i.e.}, an alternative non-Riemannian volume form 
on the space-time manifold defined in terms of an auxiliary antisymmetric gauge
field of maximal rank. The originally proposed modified-measure gravity-matter theories
\ct{TMT-orig-1}-\ct{TMT-recent-2} contained two terms in the pertinent Lagrangian action
-- one with a non-Riemannian integration measure and a second one with the
standard Riemannian integration measure (in terms of the square-root of the
determinant of the Riemannian space-time metric). An important feature was the
requirement for global Weyl-scale invariance which subsequently underwent
dynamical spontaneous breaking \ct{TMT-orig-1}. The second action term
with the standard Riemannian integration measure  might also contain a
Weyl-scale symmetry preserving $R^2$-term \ct{TMT-orig-3}.

The latter formalism yields various new interesting results
in all types of known generally covariant theories: $D=4$-dimensional models of gravity and matter fields containing 
the new measure of integration appear to be promising candidates for resolution 
of the dark energy and dark matter problems, the fifth force problem, 
and a natural mechanism for spontaneous breakdown of global Weyl-scale symmetry
\ct{TMT-orig-1}-\ct{TMT-recent-2}. Study of reparametrization invariant theories of extended objects 
(strings and branes) based on employing of a modified non-Riemannian 
world-sheet/world-volume integration measure \ct{mstring} leads to dynamically 
induced variable string/brane tension and to string models of non-abelian 
confinement, interesting consequences from the modified measures spectrum \ct{mstringspectrum}, and construction of new braneworld scenarios \ct{mstringbranes}. 
Recently \ct{nishino-rajpoot} this formalism was generalized to
the case of string and brane models in curved supergravity background. An important result for cosmology of the dynamical tension string theories is the avoidance of swampland constraints \ct{noswamplandconstraints}.

In this paper we will study a quintessential scenario where we will be driven from inflation to
an early DE phase, which then decays to the final late DE phase, through a bubble nucleation, which generalizes the model of Niedermann et. al \ct{Niedermann:2020dwg} by the use of a scale invariant two field model, where the bubble nucleation is triggered by a potential that depends on the density of the matter instead of another scalar field as in Ref.~\ct{Niedermann:2020dwg}. 

Multifield inflation has been studied by several authors see for example \ct{multifield1, multifield2, multifield3}. In the context of modified measures formalism, the ratio of two measures can become an additional scalar field if we use the second order formalism \ct{multifieldwithTMTinsecondorder}, in the present paper we will consider only the first order formulation however, and the measure field remain non dynamical, determined by a constraint and therefore they do not introduce new degrees of freedom.
Introducing two fields gives rise to very interesting  new possibilities. This is also the case when we consider multi field scale 
invariant inflationary models leading to DE/DM for the late universe, where interesting new features appear for both the inflationary phase and for the 
 DE/DM late universe phase.
In particular we will see that  the late universe acquires a fine structure with two possible vacua for the late universe that can take place at different times in the late evolution of the universe. Furthermore, in the presence of dust, the scalar field potential depends on the dust density due to the scale invariant coupling of the scalar field to the dust particles.

An interesting aspect, previously explored in Ref.~\cite{Guendelman:2022cop}—where the model under consideration was also studied—is the identification of three nearly flat regions in the scalar field potential, corresponding to inflation, early dark energy, and late-time dark energy. However, the transition between the early and late dark energy plateaus was not addressed in that work; this gap will be investigated in the present study. In Ref.~\cite{Guendelman:2022cop}, we focused instead on the dynamics of slow-roll inflationary solutions occurring on the highest plateau, and examined which of these solutions decay into the intermediate-energy plateau rather than directly into the lowest-energy (late dark energy) region. This behavior constitutes a necessary condition for realizing a NEDE scenario within our framework.

 Here we do not attempt to couple the scalar field to electromagnetism, because this will generically lead to explicit violation of scale invariance and the coupling to dust seems to achieve the desired goals already, so such a generalization does not seem to be needed. As opposed to the $\Lambda$CDM in our model DM and DE interact in  the early Universe after Inflation, when the system settles into its ground state, such interaction disappears.

 This scalar field potential has a barrier  between the Early Dark Energy and the Late Dark Energy regions of the scalar field potential,  but this barrier depends on the dust density and as the dust density dilutes,  there is a  redshift where nucleation of late dark energy bubbles in the midst of the early dark energy filled space becomes possible, and this can get us to a percolation regime, where the bubbles of the late DE sector fill up all the space, a process  which is studied in details.  The calculation of $H_0$ from early universe and CMB data in our model shows agreement with the direct redshift supernova measurements of $H_0$, so that this effect can alleviate  the $H_0$ tension.

We organize our paper as follows: In Section \ref{TMMT} we give a brief review of gravity matter formalism  with two independent non-Riemannian volume-forms. In Section \ref{flat-regions}, we describe the three infinitely large flat regions  associated to the effective potential. In Section \ref{evolution} we study the dynamics and evolution of the EDE and DM in the Einstein frame. Also, we discuss the masses of particle in the different vacua and the geodesic motion. In Section \ref{transition} we analyze the transition to late dark energy from early dark energy by tunneling. Here we determine the tunneling rate per unit volume together with the percolation parameter. In Section \ref{transition} we study the dynamics of our model related to the Friedmann equation before and after of the phase  transition. Here we find different conditions associated to the density parameters. Besides, we determine from the observational data the best-fit parameters and the different constraints on the model parameters. Finally, in Section \ref{discuss} we discuss our results. We chose units in which  $c=\hbar=1$.

\section{The origin of vacuum energy density from the standard measure of integration in General Relativity and other Theories and the idea of Non Gravitating vacuum energy}
 We are motivated in this paper by the $H_0$ problem, but before that, even a more basic and older problem, the cosmological constant problem has to be addressed and then we try to see if these two problems could be treated in an unified framework. On the mathematical level the cosmological constant problem  boils down to an asymmetry between the matter sector and the gravitational sectors of the theory concerning the role of an origin for the energy density. As it is well known, in non-gravitational physics, like in particle mechanics, for example,  the origin from which we measure energy is not important. In mathematical terms that means that the equations of motion are invariant under addition of a constant to the matter Lagrangian~$\mathcal{L}_m$
\[\mathcal{L}_m \longrightarrow  \mathcal{L}_m+C \:.\]
However, when the Lagrangian density  $\mathcal{L}_m$ is integrated with the standard Riemannian measure of integration, the square root of the determinant of the metric,  the equations of motion get an  extra contribution of the form~$C g_{\mu\nu}$  after the above shift. This implies the choice of measure of integration has a crucial effect on the cosmological constant problem, and in particular that for the standard measure of integration, the vacuum energy gravitates. This motivates us to search for alternative measures of integrations, or what is the same alternative volume forms, as we discuss next where  a shift of the Lagrangian is a symmetry. This could be achieved if the measure is a total derivative, then the shift above would not change the equations of motion of the theory and we could talk then, at least in some sense of a non gravitating vacuum energy model, as we will see, this possibility is present when we go on and discuss the Non-Riemannian Volume-Form Formalism.

\section{Consequences of the Non-Riemannian Volume-Form Formalism}
\label{consequencesofMODM}

A broad class of actively developed modified/extended gravitational theories is based on
employing alternative non-Riemannian spacetime volume-forms (metric-independent 
generally covariant volume elements) in the pertinent Lagrangian actions instead 
of, or alongside with, the canonical Riemannian volume element
given by the square-root of the determinant of the Riemannian metric. As we mentioned in the introduction, this leads to a new class of gravity-matter theories based on the 
idea that the action integral may contain a new metric-independent generally-covariant 
integration measure density, \textsl{i.e.}, an alternative non-Riemannian volume form 
on the space-time manifold defined in terms of an auxiliary antisymmetric gauge
field of maximal rank. The originally proposed modified-measure gravity-matter theories
\ct{TMT-orig-1}-\ct{TMT-recent-2} contained two terms in the pertinent Lagrangian action
-- one with a non-Riemannian integration measure and a second one with the
standard Riemannian integration measure (in terms of the square-root of the
determinant of the Riemannian space-time metric).

Volume-forms are fairly basic objects in differential geometry -- they exist
on arbitrary differentiable manifolds and define covariant (under general
coordinate reparametrizations) integration measures. It is important to
stress that the existence of volume-forms is {\em completely independent} of the
presence or absence of additional geometric structures on the manifold  -- 
Volume forms are defined \ct{spivak} by nonsingular maximal rank differential forms $\om$:
\br
\int_{\cM} \om \bigl(\ldots\bigr) = \int_{\cM} dx^D\, \Omega \bigl(\ldots\bigr)
\;\; ,
\nonu \\
\om = \frac{1}{D!}\om_{\m_1 \ldots \m_D} dx^{\m_1}\wedge \ldots \wedge dx^{\m_D}\; ,
\lab{omega-1} \\
\om_{\m_1 \ldots \m_D} = - \vareps_{\m_1 \ldots \m_D} \Omega \; ,
\nonu
\er
(our conventions for the alternating symbols $\vareps^{\m_1,\ldots,\m_D}$ and
$\vareps_{\m_1,\ldots,\m_D}$ are: $\vareps^{01\ldots D-1}=1$ and
$\vareps_{01\ldots D-1}=-1$).
The volume element density 
$\Omega$ transforms as scalar density under general coordinate reparametrizations. Notice also that 
$\Omega$  is a total derivative as well, therefore adding a constant to any Lagrangian that multiplies $\Omega$  will give rise to a total derivative that will not give a modification of the equations of motion and no generation of a cosmological constant, according to our expressed goal to have some sense of a non gravitating vacuum energy principle. 

In standard generally-covariant theories (with action $S=\int d^D\! x \sqrt{-g} \cL$)
the Riemannian spacetime volume-form is defined through the ``D-bein''
(frame-bundle) canonical one-forms $e^A = e^A_\m dx^\m$ ($A=0,\ldots ,D-1$):
\br
\om = e^0 \wedge \ldots \wedge e^{D-1} = \det\Vert e^A_\m \Vert\,
dx^{\m_1}\wedge \ldots \wedge dx^{\m_D} 
\longrightarrow \quad
\Omega = \det\Vert e^A_\m \Vert\, d^D x = \sqrt{-\det\Vert g_{\m\n}\Vert}\, d^D x \; .
\lab{omega-riemannian}
\er

Instead of, or alongside with,  $\sqrt{-g}$ we can employ one or several different  
alternative {\em non-Riemannian} volume elements as in \rf{omega-1} given by non-singular 
{\em exact} $D$-forms $\om^{(j)} = d B^{(j)}$ where:
\br
B^{(j)} = \frac{1}{(D-1)!} B^{(j)}_{\m_1\ldots\m_{D-1}}
dx^{\m_1}\wedge\ldots\wedge dx^{\m_{-1}} 
\longrightarrow \quad  \Omega^{(j)} \equiv \Phi(B^{(j)}) =
\frac{1}{(D-1)!}\vareps^{\m_1\ldots\m_D}\, \pa_{\m_1} B^{(j)}_{\m_2\ldots\m_D} \; .
\lab{Phi-D}
\er
In other words, the non-Riemannian volume elements are defined in terms of
the dual field-strengths of auxiliary rank $D-1$ tensor gauge fields 
$B^{(j)}_{\m_1\ldots\m_{D-1}}$. 

Let us again strongly emphasize that the term
``non-Riemannian'' concerns only the nature of the non-canonical volume
elements, which exist on the spacetime manifold with a standard Riemannian
geometric structure, 
torsionless affine connection $\G^\l_{\m\n}$ either independent of $g_{\m\n}$
(first-order metric-affine / Einstein-Palatini formalism) or as a Levi-Civita connection 
w.r.t. $g_{\m\n}$ (second-order purely metric / Einstein-Hilbert formalism).

The generic form of modified gravity actions involving (one or more) non-Riemannian 
volume-elements, called for short 
actions, read (henceforth $D=4$, and we will use units with 
$16\pi G_{Newton} =1$):
 {
\br
S = \int d^4 x \,\P(B^{(1)}) \bigl( R +\cL^{(1)}\bigr) + \int d^4 x \,\sum_{j\geq 2} \P(B^{(j)})\, \cL^{(j)} + \int d^4 x \,\sqrt{-g}\cL^{(0)} \; ,
\lab{NRVF-0}
\er}
where $R$ is the scalar curvature. 
The equations of motion of \rf{NRVF-0} w.r.t. the auxiliary tensor gauge fields 
$B^{(j)}_{\m\n\k}$ according to \rf{Phi-D} imply:
\br
\pa_\m \bigl( R+\cL^{(1)}\bigr)=0 \;\; ,\;\; 
\pa_\m \cL^{(j)} = 0 \;\; (j\geq 2) \; ,\, \longrightarrow \; \; R+\cL^{(1)}=M_1 \quad,\quad \cL^{(j)}= M_j  \; ,
\lab{L-M}
\er
where all $M_j$ ($j\geq 1$) are {\em free integration constants} not present in the 
original NRVF gravity action \rf{NRVF-0}. These constants of integrations play very important roles, like, if the theory possess global scale invariance, some of these constants could be responsible for the spontaneous symmetry breaking of the scale invariance and they can also add possible new vacuum states whose energy densities depend on these constants of integration, for example in our case, this will be the vacuum supporting the inflationary phase.

\section{Gravity-Matter Formalism With Two Independent Non-Riemannian Volume-Forms}
\label{TMMT}
In this section, we will present the  gravity-matter system described 
by an action which will be used
that has two independent non-Riemannian integration
measure densities defined  in the framework explained in the previous section by
\ct{quintess}
\be
S = \int d^4 x\,\P_1 (A) \Bigl\lb \frac{R}{2\kappa} + L^{(1)} \Bigr\rb +  
\int d^4 x\,\P_2 (B) \Bigl\lb L^{(2)} + \eps_S R^2 +
\frac{\P (H)}{\sqrt{-g}}\Bigr\rb \; ,
\lab{TMMT}
\ee
where $\kappa= 8\pi G = M_{P}^{-2}$ with $M_{P}$ the Planck mass and the functions $\P_{1}(A)$ and $\P_2 (B)$ correspond to  two independent non-Riemannian volume-forms
defined as 
$\P_1 (A) = \frac{1}{3!}\vareps^{\m\n\k\l} \pa_\m A_{\n\k\l} \,\,\mbox{and}\,\,\quad
\P_2 (B) = \frac{1}{3!}\vareps^{\m\n\k\l} \pa_\m B_{\n\k\l} ,$
respectively. Here we mention that
the quantities $\P_{1,2}$ take over the role of the standard Riemannian integration measure density given by 
$\sqrt{-g} \equiv \sqrt{-\det\Vert g_{\m\n}\Vert}$ and these functions  can be written   in terms of  the 
metric $g_{\m\n}$ \ct{quintess}.

In relation to the function $R = g^{\m\n} R_{\m\n}(\G)$ and the quantity $R_{\m\n}(\G)$, these denote the scalar curvature and the Ricci tensor in the first-order (Palatini) formalism, in which the affine
connection $\G^\m_{\n\l}$  \textsl{a priori}  does not dependent on the metric $g_{\m\n}$.
In addition, we have included  in this action a $R^2$-term (the Palatini form) coupled with a parameter $\epsilon_S$. We mention that $R+R^2$ action within the second order formalism was originally introduced  by Starobinsky in Ref.~\ct{starobinsky} in the context of an inflationary stage. The inflationary phase can be obtained 
 from a simple mechanism involving  the constants of integration $M_1$ and $M_2$ even  without involving an $\epsilon_S$ contribution, which would be the analogous of the Starobinsky mechanism, therefore since the constants of integration $M_1$ and $M_2$ can produce this without introducing higher powers in curvature. Then, we work with a simpler theory that produces all the necessary ingredients by taking $\epsilon_S=0$. 

Besides, the quantities $L^{(1,2)}$ correspond to two different Lagrangians associated to two  scalar matter fields and the electromagnetic field denoted by $\vp_1$ ,  $\vp_2$ and 
$A_\mu$ similarly as in  Ref.~\ct{TMT-orig-1}. In this form, the Lagrangians  $L^{(1,2)}$
 are defined by the expressions
\be
L^{(1)} = -\h g^{\m\n} \pa_\m \vp_1 \pa_\n \vp_1 -\h g^{\m\n} \pa_\m \vp_2 \pa_\n \vp_2- V(\vp_1,\vp_2) \quad ,\quad
\mbox{and}\,\,\,\,\,\,\,\,\,\,\,
L^{(2)} = U(\vp_1,\vp_2) -\frac{1}{4}F_{\mu \nu}
F^{\mu \nu},
\lab{L-2}
\ee
respectively. Here the quantity $F_{\mu\nu}$ corresponds to the antisymmetric strength tensor (electromagnetic field tensor) constructed out of the 4-potential $ A_\mu$, that is,  $F_{\mu \nu} = \partial_\mu  A_\nu - \partial_\nu  A_\mu $, the quantity $V(\vp_1,\vp_2)=V$ denotes 
 to a scalar  potential associated to the scalar fields $\vp_1$ and $\vp_2$ and it is defined as 
\be
V(\vp_1,\vp_2)=f_1\,e^{-\alpha_1\vp_1}+g_1e^{-\alpha_2\vp_2},\lab{Va1}
\ee
and the another quantity $U(\vp_1,\vp_2)=U$ corresponds to a second 
scalar potential given by 
\be
U(\vp_1,\vp_2)=f_2\,e^{-2\alpha_1\vp_1}+g_2\,e^{-2\alpha_2\vp_2},\lab{Vb1}
\ee
in which  $f_1,f_2,g_1,g_2$,$\alpha_1$ and $\alpha_2$ denote  different constants or parameters. We note that the parameters  $f_1,f_2,g_1$ and $g_2$ have dimensions of $M_{P}^4$ instead the quantities $\alpha_1$ and $\alpha_2$ have dimensions of $M_{P}^{-1}$. 
Also, in the action the function $\P (H)$ corresponds to the dual field strength of a third auxiliary 3-index antisymmetric
tensor gauge field and it is defined as 
$
\P (H) = \frac{1}{3!}\vareps^{\m\n\k\l} \pa_\m H_{\n\k\l} \; , 
$ see Ref.~\ct{TMT-orig-1}.

One may think that we are putting too many free parameters into the theory, and may think that this is not but desirable, but it is easy to see that only some combinations will appear, for this one has to notice for example that a shift of $\vp_1$, like
$\vp_1 \rightarrow \vp_1 + \Delta$,  
is equivalent to redefining the coupling constants 
$f_1$ and $f_2$ 
by the corresponding transformations
$f_1 \rightarrow e^{-\alpha_1\Delta} f_1$ 
and $f_2 \rightarrow e^{-2\alpha_1\Delta} f_2$,
which means that neither of these coupling constants can be physical by itself. 
In contrast the combination $\frac{f_1^2}{f_2}$ is invariant under this shift and in fact we will see that only this combination appears in some asymptotic values of the vacuum energy density,
by an exactly analogous argument, we can see that the g dependence can be only through $\frac{g_1^2}{g_2}$, also, as we will see, these combinations are related to the values of the flat regions of an effective potential that determines vacuum energies in the early DE and in the late DE, since we look for flat regions, these must be independent of the shifts in the fields, and this then lead us to the combinations  $\frac{f_1^2}{f_2}$  and $\frac{g_1^2}{g_2}$ without an explicit calculation. These ratios  significantly reduce the parameter space at the phenomenological level, mitigating concerns about parameter degeneracies.  
An explicit calculation confirms this as we will see.  

In relation to the scalar potentials $V$ and $U$ these have been chosen the form that the  action given 
Eq.~\rf{TMMT} becomes invariant under global Weyl-scale transformations defined as
\br
g_{\m\n} \to \l g_{\m\n} \;\; ,\;\; \G^\m_{\n\l} \to \G^\m_{\n\l} \;\; ,\;\; 
 \vp_1 \to \vp_1+\frac{1}{\alpha_1}\ln\lambda\;,\,\,\,\,\vp_2 \to \vp_2+\frac{1}{\alpha_2}\ln\lambda,
\nonu \\
A_{\m\n\k} \to \l A_{\m\n\k} \;\; ,\;\; B_{\m\n\k} \to \l^2 B_{\m\n\k}
\;\; ,\;\; H_{\m\n\k} \to H_{\m\n\k}  \;\; ,\;\; \; A_\mu  \to A_\mu,  \;\; ,\;\; \; F_{\mu \nu} \to F_{\mu \nu},
\lab{scale-transf}
\er
with $\lambda$ a constant $ F_{\mu \nu}$ is the standard gauge invariant  electromagnetic field strength defined before. Besides, we note that the difference between 
 $\alpha_1\vp_1-\alpha_2\vp_2 \to \alpha_1\vp_1-\alpha_2\vp_2,$ is invariant from the transformations defined by Eq.~\rf{scale-transf}.

Then, the equations of motion resulting  from the
variation of the action given by Eq.~\rf{TMMT} with respect to the affine connection $\G^\m_{\n\l}$ can be written as  
\be
\int d^4\,x\,\sqrt{-g} g^{\m\n} \Bigl(\frac{\P_1}{\sqrt{-g}} \Bigr) \(\nabla_\k \d\G^\k_{\m\n}
- \nabla_\m \d\G^\k_{\k\n}\) = 0 ,
\lab{var-G}
\ee
in which the quantity 
$\G^\m_{\n\l}$ represents a Levi-Civita connection defined in terms of the metric tensor as
$
\G^\m_{\n\l} = \G^\m_{\n\l}({\bar g}) = 
 {\bar g}^{\m\k}\(\pa_\n {\bar g}_{\l\k} + \pa_\l {\bar g}_{\n\k} 
- \pa_\k {\bar g}_{\n\l}\)/2$,
w.r.t. to the Weyl-rescaled metric ${\bar g}_{\m\n}$, such that
\be
{\bar g}_{\m\n} = \chi_1 \; g_{\m\n} \;\; ,\;\; \mbox{and}\,\,\,\,
\chi_1 \equiv \frac{\P_1 (A)}{\sqrt{-g}} \;.
\lab{bar-g}
\ee

Moreover, considering the variation of the action  defined  by Eq.~\rf{TMMT} with respect to the  auxiliary tensor gauge fields
$A_{\m\n\l}$, $B_{\m\n\l}$ and $H_{\m\n\l}$ we find  the equations
\be
\pa_\m \Bigl\lb \frac{R}{2\kappa} + L^{(1)} \Bigr\rb = 0 \quad, \quad
\pa_\m \Bigl\lb L^{(2)} +  \frac{\P (H)}{\sqrt{-g}}\Bigr\rb = 0 
\quad\,\,\,\,\mbox{and}\,\,\,\, \quad \pa_\m \Bigl(\frac{\P_2 (B)}{\sqrt{-g}}\Bigr) = 0 \; ,
\lab{A-B-H-eqs}
\ee
respectively. The  solutions of Eq.~\rf{A-B-H-eqs} can be written as
\be
\frac{\P_2 (B)}{\sqrt{-g}} \equiv \chi_2   ,\;\;\;\,\,\,\,\,\,\,\,
\frac{R}{2\kappa} + L^{(1)} = - M_1   \,\,\;\;\;\,\mbox{and}\,\,\,\,\,\,\,\,
L^{(2)} +  \frac{\P (H)}{\sqrt{-g}} = - M_2  ,
\lab{integr-const}
\ee
where  $M_1$, $M_2$ and $\chi_2$ correspond to  integration constants. We mention that the constants  $M_1$ and $M_2$ are arbitrary  and with dimensions of $M_P^4$. However, the integration constant $\chi_2$ is
also arbitrary and dimensionless. 
In relation to the  constant $\chi_2$ in Eq.~\rf{integr-const}, we mention that it preserves
global Weyl-scale invariance in Eq.~\rf{scale-transf}. However, 
 the another integration constants $M_1,\, M_2$
  dynamical spontaneous breakdown of global Weyl-scale invariance 
under \rf{scale-transf} product of the scale non-invariant solutions 
obtained in  Eq.~\rf{integr-const}. 

One should point out that the integration constants $M_1$ and $M_2$ by themselves violate spontaneously the scale invariance of the theory, since while
$\frac{R}{2\kappa} + L^{(1)}$  transforms under a scale transformation, the integration of the equations of motion fix this to be equal to $- M_1 $ and in a similar way, $L^{(2)} +  \frac{\P (H)}{\sqrt{-g}}$ transforms under a scale transformation \rf{scale-transf}, the integration of the equations of motion fix this to be equal to $- M_2 $ . Notice that while $M_1 $ and $ M_2 $ would change as in $M_1 \rightarrow M_1/\lambda $ and $ M_2 \rightarrow M_2/\lambda^2$ while we perform a scale transformation, a particular combination, $\frac{M_1^2}{M_2}$ does not. This combination must represent a real physical quantity because of its scale invariant nature and we expect it to appear in physical quantities, like, as we will see, in the expectation value of the the inflationary phase of the universe. This explain a rather substantial reduction of the relevant parameters, from  $ M_1 $ and  $M_2 $  
to just $\frac{M_1^2}{M_2}$. The scale transformations \rf{scale-transf} bring us from one inflationary vacuum to another one with the same energy density, which will mean, with the same value of  $\frac{M_1^2}{M_2}$, this symmetry translates into a flat region in the effective potential which this argument anticipates.

Also, the variation of the action \rf{TMMT} w.r.t. $g_{\m\n}$ and considering the quantities defined by Eq.~\rf{integr-const} we find the expression
\be
\chi_1 \Bigl\lb R_{\m\n} + \h\( g_{\m\n}L^{(1)} - T^{(1)}_{\m\n}\)\Bigr\rb -
\h \chi_2 \Bigl\lb T^{(2)}_{\m\n} + g_{\m\n} \; M_2
\Bigr\rb = 0 \; ,
\lab{pre-einstein-eqs}
\ee
where  the quantities $T^{(1,2)}_{\m\n}$ denote  the energy-momentum tensors associated to  the scalar
field Lagrangians defined by  the standard expressions 
\be
T^{(1,2)}_{\m\n} = g_{\m\n} L^{(1,2)} - 2 \partder{}{g^{\m\n}} L^{(1,2)} \; .
\lab{EM-tensor}
\ee

On the other hand, taking the trace of Eq.~\rf{pre-einstein-eqs} and considering the second term of 
Eq.~\rf{integr-const}, we obtain that the scale factor $\chi_1$ is given by
\be
\chi_1 = 2 \chi_2 \frac{T^{(2)}/4 + M_2}{L^{(1)} - T^{(1)}/2 - M_1} \; ,
\lab{chi-1}
\ee
where the quantities $T^{(1,2)} = g^{\m\n} T^{(1,2)}_{\m\n}$. 

Now, by considering the second term of Eq.~\rf{integr-const} and combining with the  Eq.~\rf{pre-einstein-eqs}, we find   
 the Einstein-like equations given by 
\br
R_{\m\n} - \h g_{\m\n}R = 2\kappa\left(\h g_{\m\n}\(L^{(1)} + M_1\)
+ \frac{1}{2}\(T^{(1)}_{\m\n} - g_{\m\n}L^{(1)}\)
+ \frac{\chi_2}{2\chi_1 } \Bigl\lb T^{(2)}_{\m\n} + 
g_{\m\n}\,M_2 \Bigr\rb \right) \; .
\lab{einstein-like-eqs}
\er

However, we can write Eq.\rf{einstein-like-eqs} in the standard form of Einstein 
equations  
$R_{\m\n}({\bar g}) - \h {\bar g}_{\m\n} R({\bar g}) = \kappa\, T^{\rm eff}_{\m\n},$
where the energy-momentum tensor  $T^{\rm eff}_{\m\n}$ is defined as (similarly to \rf{EM-tensor})
$T^{\rm eff}_{\m\n} = g_{\m\n} L_{\rm eff} - 2 \partder{}{g^{\m\n}} L_{\rm eff} ,$
and the  effective scalar field Lagrangian  in the Einstein-frame can be written as 
\be
L_{\rm eff} = \frac{1}{\chi_1}\Bigl\{ L^{(1)} + M_1 +
\frac{\chi_2}{\chi_1}\Bigl\lb \bar{L}^{(2)} + M_2  
\Bigr\rb\Bigr\} \;  -\frac{1}{4}F_{\mu \nu}
F^{\mu \nu},
\lab{L-eff}
\ee
in which the quantities  $L^{(1}, \bar{L}^{(2)}$ correspond to the Lagrangian densities given by 
$L^{(1)} = \chi_1\, (X_1+X_2) - V$ and  $\bar{L}^{(2)} =  U .$
Notice that we treat now the electromagnetic contribution separately, because of the conformal invariance of this term, so that the electromagnetic contribution is the same in any frame, for this reason also, we do not include 
$-\frac{1}{4}F_{\mu \nu}F^{\mu \nu}$ in $\bar{L}^{(2)}$ and instead it appears as a different contribution in \rf{L-eff}.
Here we have considered the short-hand notation for the
 kinetic terms $X_1$ and $X_2$ associated to the scalar fields $\vp_1$ and $\vp_2$ defined as
\be
 X_1 \equiv - \h {\bar g}^{\m\n}\pa_\m \vp_1 \pa_\n \vp_1,\,\,\,\,\mbox{and}\,\,\,\,\,\,\, X_2 \equiv - \h {\bar g}^{\m\n}\pa_\m \vp_2 \pa_\n \vp_2.
\lab{X-def}
\ee

Now, from  Eq.~\rf{chi-1}  and considering  $L^{(1)}$ and $L^{(1)}$
we find that the function $\chi_1$ results

\be
\chi_1 = \frac{2\chi_2\Bigl\lb U+M_2 \Bigr\rb}{(V-M_1)}
\, \; .
\lab{chi-Omega}
\ee
Thus, combining Eqs.~\rf{L-eff} and \rf{chi-Omega}, we obtain that  the Lagrangian $L_{\rm eff}$ relative to the two scalar fields $\vp_1$ and $\vp_2$, in the framework of the the Einstein can be written as
\be
L_{\rm eff} =  X_1+X_2  - U_{\rm eff}(\vp_1,\vp_2) \;  -\frac{1}{4}F_{\mu \nu}
F^{\mu \nu}.
\lab{L-eff-final}
\ee
Now above, just as we did when defining the kinetic terms $X_1 $, $X_2 $,  we raise indices, now to define for example $F^{\mu \nu}$, we do it  with the inverse of the metric in the Einstein frame ${\bar g}^{\m\n}$. Also,  the effective scalar  potential $U_{\rm eff}(\vp_1,\vp_2)$ associated to the scalar fields $\vp_1$ and $\vp_2$ yields
\be
U_{\rm eff} (\vp_1,\vp_2) = 
\frac{(V - M_1)^2}{4\chi_2 \Bigl\lb U + M_2 \Bigr\rb}
= \frac{\(f_1 e^{-\a_1\vp_1}+g_1e^{-\a_2\vp_2}-M_1\)^2}{4\chi_2\,\Bigl\lb 
f_2 e^{-2\a_1\vp_1}+g_2 e^{-2\a_2\vp_2}+ M_2 \Bigr\rb} \; .
\lab{U-eff} 
\ee
Here we have utilized  the  scalar potentials $V$ and $U$ defined by  Eqs.~\rf{Va1} and \rf{Vb1}, respectively.

It is important to emphasize that the functional form of the scalar fields potentials $V, U$ and $U_{\rm eff}$, considered in this study  is not chosen arbitrarily. Rather, it arises from the requirement that the underlying gravity matter action be invariant under global Weyl-scale transformations. In this way, this symmetry places  strong restrictions on the allowed functional forms to exponential potentials $V$, $U$ and $U_{\rm eff}$ and then,  the structure of the potentials is determined by the symmetry of the theory itself.

\section{  Effective Scalar Potential: Flat regions}
\label{flat-regions}

From the effective potential $U_{\rm eff}$ given by Eq.~\rf{U-eff}, we can note that the presence of three infinitely large flat regions. These regions can be obtained considering different  large positive values of the  fields $\vp_1$ and $\vp_2$, respectively.
Thus, for the case in which we assume  large positive values of the fields $\vp_1$ and $\vp_2$, we find  that the effective potential is reduced to
\br
U_{\rm eff}(\vp_1,\vp_2) \simeq U_{(\vp_1\to+\infty,\vp_2\to+\infty)}= U_{(++)} = 
\frac{M_1^2}{4\chi_2\,M_2} \; .
\lab{U-minus} 
\er
In the situation in which  we only consider a  large negative for the scalar field $\vp_1$, the effective potential can be associated  to  another  flat region defined by 
\br
U_{\rm eff}(\vp_1,\vp_2) \simeq U_{(\vp_1\to-\infty)} \equiv 
\frac{f_1^2}{4\chi_2 \,f_2} \; .
\lab{U-plus} 
\er
In the case in which we only assume a large negative for the scalar field $\vp_2$, we have
\br
U_{\rm eff}(\vp_1,\vp_2) \simeq U_{(\vp_2\to-\infty)} \equiv 
\frac{g_1^2}{4\chi_2 \,g_2} \; .
\lab{U-plus2} 
\er

In relation to the three flat regions \rf{U-minus},   \rf{U-plus} and \rf{U-plus2}, we can assume that these regions can be associated  
to the evolution of the early and the late universe, respectively. Specifically,  we can consider that the first flat region can be related to the inflationary epoch, the second flat region to the early dark energy and the third region can be associated to the present dark energy. Under energy considerations, we can infer that 
 the ratio of the coupling constants associated to the flat regions  during the different epochs satisfy
\be
\frac{M_1^2}{M_2} \gg \frac{f_1^2}{f_2} > \frac{g_1^2}{g_2}.
\lab{early-vs-late}
\ee
Thus, from Eq.~\rf{early-vs-late}, we ensure that the vacuum energy density during the inflationary scenario $U_{(++)}$ is much bigger
than both the early dark energy and  the current dark energy.

Additionally, considering the cosmological perturbations, described by the tensor-to-scalar ratio $r$ and the scalar power perturbation $\cP_S$,
we can estimate that the first flat region of the effective potential associated with the inflationary epoch results in 
$\kappa^2\,U_{(++)} \sim \kappa^2M_1^2/\chi_2M_2\sim 6\pi^2\,r\,\cP_S \sim 10^{-8}\,$, see Refs.~\ct{Planck1,Planck2,Lplanck}.

\section{Dark energy and dark matter epochs}
\label{evolution}

In this section we will study the dynamics and 
 the evolution of the early dark energy and dark matter.
During the evolution of the universe, a phase of particle creation is necessary to produce both dark matter and ordinary matter. This particle production can occur through various mechanisms, even in scenarios where a single scalar field is coupled to different energy measures
\ct{reheatingwithtwomeasures}. In this sense, we can incorporate  a dark matter particles contribution, under a scale invariant form given by the matter action $S_m$ specified by
\begin{equation}
\label{particles}\\
S_{m}=\int( \Phi_1 +b_{m}e^{\kappa_1\p_2} \sqrt{-g})\,L_m d^{4}x ,
\end{equation}
 where the quantity $b_{m}$ corresponds to a constant that accounts for  the strength of the coupling between the scalar field $\p_2$ and the term $\sqrt{-\bar{g}}$ in the Einstein frame.
Here the scalar field $\phi_2$ is introduced from a scalar transformation in terms of the original fields $\varphi_1$ and $\varphi_2$  as  with the field $\phi_1$ \ct{reheatingwithtwomeasures}
 \begin{equation}  
 \phi_1=\frac{\alpha_1\varphi_1-\alpha_2\varphi_2}{\sqrt{\alpha_1^2+\alpha_2^2}},\,\,\,\,\,\,\mbox{and}\,\,\,\,\,\,\,
 \phi_2=\frac{\alpha_2\varphi_1+\alpha_1\varphi_2}{\sqrt{\alpha_1^2+\alpha_2^2}},\label{nF}
  \end{equation} 
with which this transformation is orthogonal,  $\dot{\phi_1}^2+\dot{\phi_2}^2=\dot{\varphi_1}^2+\dot{\varphi_2}^2$.

Besides, the matter Lagrangian density  $L_m$ is defined by
 \begin{equation}  \label{particles1}
L_m=-\sum_{i} m_{i}\int e^{\kappa_2\p_2}
\sqrt{-g_{\alpha\beta}\frac{dx_i^{\alpha}}{d\lambda}\frac{dx_i^{\beta}}{d\lambda}}\,
\frac{\delta^4(x-x_i(\lambda))}{\sqrt{-g}}d\lambda ,
\end{equation}
in which the quantities $\kappa_1$ and $\kappa_2$ into Eqs.~(\ref{particles}) and (\ref{particles1}) are constants and these satisfy the condition of scale invariance.  In relation to the invariance, this condition  determines that the coupling constants to be equal to $\kappa_1 = -\frac{\alpha_1\a_2}{\sqrt{\a_1^2+\a_2^2}}$ and $\kappa_2 =-\frac{1}{2}\kappa_1$, respectively \cite{Guendelman:2022cop}. Also, 
the quantity $m_i$ is the mass parameter of the ``{\it{i-th}}" particle associated to the matter.

By assuming these conditions, the existence  of matter induces a potential related to  the scalar field $\p_2$ since there is a scalar field dependence $\p_2$. Thus, 
the scalar field $\p_2$  considering the dust particles are co-moving, the energy density associated to the matter can be written as
\begin{equation} 
 \rho_m=(e^{-\frac{1}{2}\kappa_1\p_2}\Phi_1 + b_{m}e^{\frac{1}{2}\kappa_1\p_2} \sqrt{-g}  )  n\,,
 \lab{matter potential cosmological}
\end{equation}
in which $n$ corresponds to the mass density of the dust in the original framework and this density  is  diluted  proportionally to $\frac{1}{a^3}$. 
Precisely, the mass density is defined as $n = \sum_{i} m_{i} \delta^3(x-x_i(\lambda)) \frac{1}{a^3}$. 
This is due to the fact that all the temporal components of the particles are equal to the cosmic time. Performing the $\lambda$ integration, which sets $\lambda = t$ and thus $\frac{dx_i^{0}}{d\lambda} = 1$, the square root of the temporal component of the metric in both the numerator and denominator of (\ref{particles1}) cancels out, leaving us with a factor of $\frac{1}{a^3}$.

Following Ref.~\cite{Guendelman:2022cop}, we can consider that this energy density is extremized by the condition

\begin{equation}
\Phi_1 - b_{m}e^{\kappa_1\p_2} \sqrt{-g} = 0\,.\label{ec1q}
\end{equation}
In addition, we comment that this condition also eliminates all forms of non-canonical anomalous effects, such as the appearance of pressure in the contribution to the energy-momentum related to the different particles. Moreover, 
we mention that the scale factor $a$ corresponds to the original frame and not in the Einstein frame in which the scale factor corresponds to $\bar{a}$. Here the relation for the scale factor in both frames is defined as follows 
$${\bar a} =(\chi_1)^{\frac{1}{2}} a.$$
Thus, expressing then the energy density associated to the matter given by  \rf{matter potential cosmological} in Einstein frame, considering that the mass density is $n  = \frac{c}{a^3} $, then the energy density $\rho_m$ in the Einstein frame can be written as 
\begin{equation} 
 \rho_m=\left(e^{-\frac{1}{2}\kappa_1\p_2}(\chi_1)^{\frac{1}{2}} +  b_{m}e^{\frac{1}{2}\kappa_1\p_2} (\chi_1)^{-\frac{1}{2}} \right) \frac{c}{{\bar a}^3},
 \lab{matter potential cosmological EINSTEIN}
\end{equation}
where $c$ denotes a constant.

Independently of that defining $ F = e^{-\frac{1}{2}\kappa_1\p_2}(\chi_1)^{\frac{1}{2}}$,
the form of Eq.\rf{matter potential cosmological EINSTEIN} given by $F +  b_{m} F^{-1}  $ is extremized at 
\begin{equation}
(1 -b_{m} F^{-2})F^{'}= 0 ,\label{FS2}
\end{equation}
where $F^{'}$ represents derivative with respect to any of the fields.
From Eq.~(\ref{FS2}), we note that there is a solution given by 
\begin{equation}
b_{m} F^{-2}= 1. \label{FS1}
\end{equation}
However, there could be another solution if  $F$ itself is extremized, i.e $F^{'}= 0 $.

\begin{figure}[t]
\centering
\subfigure{\includegraphics[scale=1.0]{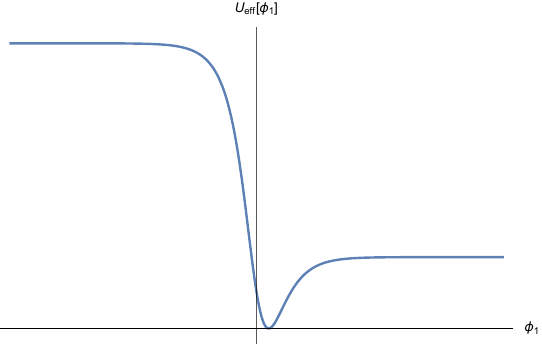}}
\caption{Schematic representation the effective potential $U_{\rm eff}(\phi_1)$  as a function of the scalar field $\phi_1$.}
\label{fig:evolucion1-1}
\end{figure}

Let us see now that  the function $ F = e^{-\frac{1}{2}\kappa_1\p_2}(\chi_1)^{\frac{1}{2}}=F(\phi_1)$ is only a function of $\p_1$, see Eq.~(\ref{nF}). To simplify matters, let us calculate $F^2$, $F^2 =  e^{-\kappa_1\p_2 }\chi_1$, to start with, let us express $\chi_1$ as the product of a scale invariant function, the effective potential in the absence of matter times an additional function, whose $\p_2$ dependence exactly cancels that of $e^{-\kappa_1\p_2 }$. According to \rf{chi-Omega}, and neglecting the integration constants $M_1$ and $M_2$, we have 
  \be
\chi_1 = \frac{2\chi_2\Bigl\lb U \Bigr\rb}{(V)} =\frac{2\chi_2\Bigl\lb U \Bigr\rb}{(V)} = 2\frac{V} {U_{eff}}
\, \; .
\lab{chi-Omega con Ueef como factor}
\ee
In the case we neglect the constants of integration, the effective potential $U_{eff}$ depends only on $\p_1$, since
using  Eq.~\rf{U-eff} and considering  the region in which $f_1e^{-\a_1\vp_1}+g_1e^{-\a_2\vp_2}\gg M_1$ and $f_2e^{-2\a_1\vp_1}+g_2e^{-2\a_2\vp_2}\gg M_2$, the effective potential reduces to
\be
U_{eff}(\vp_1,\vp_2)=\frac{(f_1e^{-\a_1\,\vp_1}+g_1e^{-\a_2\vp_2})^2}{4\chi_2(f_2e^{-2\a_1\vp_1}+g_2e^{-2\a_2\vp_2})},\lab{U4}
\ee
and from Eq.~(\ref{nF}) we have that the effective potential given by Eq.~\rf{U4} can be rewritten as a function of the  single scalar field $\phi_1$ results
\be
U_{eff}(\phi_1)=\frac{(f_1e^{-\sqrt{\a_1^2+\a_2^2}\,\phi_1}+g_1)^2}{4\chi_2(f_2e^{-2\sqrt{\a_1^2+\a_2^2}\phi_1}+g_2)}.\label{UUU}
\ee
In Fig.~\ref{fig:evolucion1-1}, we present the evolution of the effective potential $U_{{eff}}$ as a function of the scalar field $\phi_1$, as given by Eq.~(\ref{UUU}).
 From this plot, we observe that for large negative values of the field $\phi_1$, the effective potential exhibits a flat region approximately given by $U_{{eff}} \simeq \frac{f_1^2}{4\chi_2 f_2}$. A second flat region appears for large positive values of the scalar field, where the effective potential approaches $U_{{eff}} \simeq \frac{g_1^2}{4\chi_2 g_2}$.

Now notice that $e^{-\kappa_1\p_2 } V$ also depends only on $\p_1$,
this is because 
$$e^{-\kappa_1\p_2 } V =e^{-\kappa_1\p_2 } (f_1\,e^{-\alpha_1\vp_1}+g_1e^{-\alpha_2\vp_2}),$$ expressing $\vp_1$ and $\vp_2$ in terms of $\p_1$ and $\p_2$ from Eq.(\ref{nF}),
we obtain
$$\alpha_1\vp_1= \frac{\a_1^2\phi_1}{\sqrt{\a_1^2+\a_2^2}} + \frac{\a_1 \a_2\phi_2}{\sqrt{\a_1^2+\a_2^2}} = \frac{\a_1^2\phi_1}{\sqrt{\a_1^2+\a_2^2}} - \kappa_1\p_2,$$
and
$$\alpha_2\vp_2= \frac{\a_1 \a_2 \p_2 }{\sqrt{\a_1^2+\a_2^2}} - \frac{\a_2^2\p_1 }{\sqrt{\a_1^2+\a_2^2}}=  - \frac{\a_2^2\p_1 }{\sqrt{\a_1^2+\a_2^2}} - \kappa_1\p_2 ,$$
where we recall that $\kappa_1$ is  defined $\kappa_1=-\alpha_1\alpha_2/\sqrt{\alpha_1^2+\alpha_2^2}$. Thus, 
when inserting back into the expression for $F^{-2}$ we can see that the dependence of $\p_2 $ cancels out. So the energy density associated to the matter $\rho_m$ depends only on $\p_1 $ and the scale factor, as it should be because $\p_2 $ transforms under a scale transformation, while $\p_1 $ does not.
In this way, the final result for the function $F$ as a function of the new scalar field $\phi_1$ is given by 
  \be
F(\phi_1)=F  =  \left[2\frac{(f_1 e^{-\frac{\a_1^2 \p_1}{\sqrt{\a_1^2+\a_2^2}}} + g_1e^{\frac{\a_2^2 \p_1}{\sqrt{\a_1^2+\a_2^2}}}) } {U_{eff}(\phi_1)
}\right]^{\frac{1}{2} },
\lab{F}
\ee
with the effective potential $U_{eff}(\phi_1)$ is given by Eq.~(\ref{UUU}). In this form, using Eq.~\rf{matter potential cosmological EINSTEIN} we find that the energy density related to the matter in terms of the new scalar field $\phi_1$ and the scale factor  in the Einstein frame can be written as
\be
\rho_m(\phi_1,\bar{a})=\rho_m= \left(\left[2\frac{(f_1 e^{-\frac{\a_1^2 \p_1}{\sqrt{\a_1^2+\a_2^2}}} + g_1e^{\frac{\a_2^2 \p_1}{\sqrt{\a_1^2+\a_2^2}}}) } {U_{eff}}\right]^{\frac{1}{2} }+b_m\,\left[2\frac{(f_1 e^{-\frac{\a_1^2 \p_1}{\sqrt{\a_1^2+\a_2^2}}} + g_1e^{\frac{\a_2^2 \p_1}{\sqrt{\a_1^2+\a_2^2}}}) } {U_{eff}}\right]^{-\frac{1}{2} }\right)\,\frac{c}{\bar{a}^3}.\label{rhom2}
\ee

On the other hand, in relation to the  effects were recognized in a scale invariant two measure model of gravity  in  Ref.~\cite{5thforce} to obtain the avoidance of the Fifth Force Problem, which the scalar field $\p_2$, the ¨dilaton¨, could  cause, since it is a massless field. In this sense, the Fifth Force Problem is also avoided, and this can be ensured when the scalar field $\p_1$ adjusts itself to satisfy the Eq.~(\ref{ec1q}). Thus, we obtain   that the equation for scalar field  $\phi_1$ becomes \cite{Guendelman:2022cop}
 \be
 2\chi_2f_2e^{-\frac{\a_1^2}{\sqrt{\a_1^2+\a_2^2}}\p_1}+2\chi_2g_2e^{\frac{\a_1^2}{\sqrt{\a_1^2+\a_2^2}}\p_1}=b_mf_1+b_mg_1e^{\sqrt{\a_1^2+\a_2^2}\,\p_1}.\lab{ppp}
 \ee
 In this way, the above equation   determines the value of scalar field  $\phi_1$ to be a given constant  and then the speed  of the scalar field $\dot{\p}_1=0$. To find the value of the scalar field $\phi_1$ we can consider the change of variable  $x=e^{\frac{\a_1^2\p_1}{\sqrt{\a_1^2+\a_2^2}}}$ with which Eq.~\rf{ppp} results
 \be
 2\chi_2g_2 x^2 - b_m g_1x^{\frac{2\a_1^2+\a_2^2}{\a_1^2}} - b_m f_1x + 2\chi_2 f_2    = 0\,.\lab{pp1}
 \ee
 
 To determine a solution for the  field $\phi_1$ from Eq.~\rf{pp1}, we assume that for very large value of the scalar field $\phi_1$ (or analogously $x\rightarrow \infty$)  the dominate terms of Eq.\rf{pp1} are given by 

\be  
2\chi_2g_2 x^2 - b_m g_1x^{\frac{2\a_1^2+\a_2^2}{\a_1^2}}\sim 0,\,\,\,\mbox{with which}\,\,\,\,x\sim\left(\frac{2\chi_2g_2}{g_1 b_m}\right)^{(\a_1/\a_2)^2},\lab{x1}
\ee
where for consistency, we must choose that the ratio $(\chi_2g_2/g_1b_m)\rightarrow \infty$.
Thus, we find that  the value of the scalar field $\phi_1$ at this point becomes \cite{Guendelman:2022cop}
\be
\phi_{1_(+)}\sim\frac{\sqrt{\a_1^2+\a_2^2}}{\a_2^2}\,\ln\left[\frac{2\chi_2g_2}{f_1 b_m}\right].\lab{fp}
\ee

On the other hand, in the region in which the scalar field $\phi_1\rightarrow -\infty$ (or equivalently $x\rightarrow 0$)
we determine that the dominant terms are given by

\be  \lab{largenegativefield}
- b_m f_1x + 2\chi_2 f_2 \sim 0, \,\,\,\,\mbox{and then }\,\,\,x\sim \left(\frac{2\chi_2 f_2}{f_1 b_m}\right)\rightarrow 0,
\ee
in which the value of the scalar field $\p_1$ at this point is \cite{Guendelman:2022cop}
\be
\phi_{1_(-)}\sim\frac{\sqrt{\a_1^2+\a_2^2}}{\a_1^2}\,\ln\left[\frac{2\chi_2f_2}{f_1b_m}\right].\lab{fm}
\ee

In what follows of this section, we analysis  the dynamics of the dark energy together the dark matter characterized by the energy density $\rho_m$ defined by Eq.~(\ref{rhom2}). 

In relation to the dynamics of the universe, we can assume that the metric is described by the   flat Friedmann-Lemaitre-Robertson-Walker (FRW)
 metric  in the Einstein frame defined as \ct{early-univ}
\be
ds^2 = - d\bar{t}\,^2 + \bar{a}\,^2(\bar{t}) \Bigl\lb d\bar{r}^2
+ \bar{r}^2 (d\bar{\th}^2 + \sin^2\bar{\th} d\bar{\phi}^2)\Bigr\rb,
\lab{FLRW}
\ee
in which the quantity  $\bar{a}(\bar{t})$ corresponds to  the scale factor in the Einstein frame. 

In this way,  the dynamics of the universe described by the  Friedmann equations can be written as 
\be
\frac{\ddot{\bar{a}}}{\bar{a}}= - \frac{\kappa}{6} (\rho + 3p), \quad \quad\mbox{and}\,\,\,\,\,\,\,\,\,
\bar{H}\,^2  = \frac{\kappa}{3}\,\,\rho  ,
\lab{friedman-eqs}
\ee
where the Hubble parameter in the Einstein frame is defined as $ \bar{H}= \frac{\dot{\bar{a}}}{\bar{a}}$. In the following, we will assume that the dots denote derivatives with respect to the time $\bar{t}$ in the Einstein frame.

Besides,  the total energy density $\rho$ and the total pressure $p$ associated to the matter and the two homogeneous scalar fields $\vp_1 = \vp_1 (\bar{t})$ and $\vp_2 = \vp_2 (\bar{t})$ are defined as $\rho=\rho_{\vp_1\vp_2}+\rho_m$ and $p=p_{\vp_1\vp_2}$, respectively.
 Here the energy density and pressure related to the two scalar fields are given by 
\be
\rho_{\vp_1\vp_2} =X_1+X_2+ U_{\rm eff}(\vp_1,\vp_2)=\h  \vpdot_1^2 + \h  \vpdot_2^2+ U_{\rm eff}(\vp_1,\vp_2) \; ,
\lab{rho-def} 
\ee
and 
\be
p_{\vp_1\vp_2} = X_1+X_2-U_{\rm eff}(\vp_1,\vp_2)=\h  \vpdot_1^2 + \h  \vpdot_2^2 - U_{\rm eff}(\vp_1,\vp_2).
\lab{p-def}
\ee
Further, from Eq.~\rf{L-eff-final} we have that  the scalar
 equations of motion for the two scalar fields $\vp_1$ and $\vp_2$ are 
\be
\vpddot_1 + 3 \bar{H} \vpdot_1  + \partial U_{\rm eff}/\partial\vp_1 
 = 0 \; ,
\lab{vp-eqs-full}
\ee
and
\be
\vpddot_2 + 3 \bar{H} \vpdot_2  + \partial U_{\rm eff}/\partial\vp_2 
 = 0 \; ,
\lab{vp-eqs-full2}
\ee
respectively.

In this context, we can rewritten the  flat-Friedmann equation for this stage as
\be
\bar{H}\,^2=\frac{\kappa}{3}\Big[\frac{\dot{\vp}_1^2}{2}+\frac{\dot{\vp}_2^2}{2}+V_{T}(\p_1,\bar{a})\Big]=\frac{\kappa}{3}\Big[\frac{\dot{\p}_1^2}{2}+\frac{\dot{\p}_2^2}{2}+V_{T}(\p_1,\bar{a})\Big],
\ee
where we have used the transformation orthogonal between the scalar fields $(\vp_1,\vp_2)$ and $(\p_1,\p_2)$. Also, we have defined that 
the total effective potential $V_{T}$ as a function of the scalar field $\p_1$ and the scale factor in the Einstein frame and it is given by
\begin{equation}\label{VT}
    V_T(\phi_1, \bar{a}) = \Big[F(\phi_1) + b_m\,F^{-1}(\phi_1) \Big] \left( \frac{c}{\bar{a}^3}\right) + U_{\rm eff}(\phi_1) \,.
\end{equation}
Here the effective potential $U_{\rm eff}(\p_1)$ is given by Eq.~(\ref{UUU}) and  the function $F(\p_1)$ defined by Eq.~\rf{F} can be rewritten as 
\begin{equation}\label{F1}
    F(\phi_1) =  \left(\left| \frac{8 \chi_2 \left(f_2 e^{-2\sqrt{\alpha_1^2+\alpha_2^2}\,\phi_1} + g_2 \right)}{e^{-\frac{\alpha_2^2}{\sqrt{\alpha_1^2+\alpha_2^2}}\phi_1}\left(f_1 e^{-\sqrt{\alpha_1^2+\alpha_2^2}\,\phi_1} + g_1 \right)} \right|\right)^\frac{1}{2} ,
\end{equation}
where we have used the absolute value in the function $F$ to ensure that this function is a real quantity when the parameter $g_1<0$. In the following, we will consider the parameter $g_1$ to be a negative quantity.

\subsection{Masses of particles in the different vacua}

At the two minima of the total potential, one can calculate the masses of particles and they are the same. This is very simple to see from the fact that at the two minima the relation $b_{m} F^{-2}= 1 $
holds at the two minima, so at the two minima the value of $F$ is the same, but the value of the mass corresponds to the coefficient of the
$\frac{1}{\bar{a}^3}$ term in the total potential, which depends only on $F$ and 
since $F$ is the same at the two minima, the masses of particles are the same at the two vacua.

\subsection{Geodesic motion for point particles in TMT}
IF in  our analysis we want to also consider point particle motion with geodesic motion, i.e, that will behave like normal dust matter that will not be affected by the scalar field,  it is possible to formulate  such point particle 
model of matter in four dimensions ($D=4$) for TMT in a way
such that the modified measure of matter that couples to the matter 
as in 
\begin{equation}
S_{m geodesic} =\int\Phi_{1} L_{m geodesic}d^{D}x,
    \label{Act1}
\end{equation}
and the Lagrangian satisfies

\begin{equation}
g^{\mu \nu }\frac{\partial L_{m geodesic}}{\partial g^{\mu \nu } } -L_{m geodesic}=0,
\label{con}
\end{equation}
is satisfied, which is the statement that the Lagrangian has homogeneity 1 with respect to scalings of the  metric $g^{\mu \nu}$,
which in turns turns out to be the statement of scale invariance, with no coupling to any scalar field.
In this case the matter does not have a direct coupling to the scalar field, does not  modify the constraint that allows us to solve for the measure, the equation of the scalar field and produce geodesic motion for the point particles in TMT, 
This is because for the
free falling point
particle a variety of actions are possible (and are equivalent in the
context of general relativity).  The usual actions in the 4-dimensional 
space-time with the metric $g_{\mu\nu}$ are taken to be $S=-m\int F(y)ds$, 
where $y=g_{\mu\nu}\frac{dX^{\mu}}{ds}\frac{dX^{\nu}}{ds}$ and $s$ is
determined to be an affine parameter except if $F=\sqrt{y}$, which is
the case of reparametrization invariance.  In our model we must take
$S_{m geodesic}=-m\int L_{m geodesic}\Phi d^{4}x$ with $L_{m geodesic}=
-m\int ds\frac{\delta^{4}(x-X(s))}{\sqrt{-g}}F(y(X(s)))$ where $\int 
L_{m geodesic}\sqrt{-g}d^{4}x$ would be the action of a point particle in 4
dimensions in the usual theory.  For the choice $F=y$, constraint (\ref{con} )
 is satisfied and a geodesic equation (and therefore the equivalence principle) is satisfied in terms of the Einstein frame metric.  Unlike the case of general relativity,
different choices of $F$ lead to in-equivalent theories. For a discussion see \cite{ongeodesicsinTMT}.

\section{Transition to Late Dark Energy from Early Dark Energy by Tunneling}\label{transition}

The New Early Dark Energy model (NEDE), see Refs.~\cite{Niedermann:2019olb, Niedermann:2020dwg}, falls in the category of early time modifications of $\Lambda CDM$. It suggests a solution to the Hubble tension by means of reducing the size of the sound horizon, $r_s$. These models add a new energy component which initially behaves as dark energy up to a certain time $t'$ (redshift $z'$) at which it begins to redshift away. In order to have a noticeable impact on the Hubble parameter, it is required that the decay of this new component must occur not too long before recombination, around matter-radiation equality.
Thereafter, the energy fraction stored in it starts to decay rapidly, i.e., faster than radiation; in this way, the model avoids creating big deviations in other cosmological parameters. In particular in the NEDE models it is consider that this scheme is realized by a first order phase transition in a dark sector at zero temperature. Such a phase transition will have the effect of lowering an initially high value of the cosmological constant in the early Universe down to the value today, inferred from the measurement
of $H_0$. 

The main features distinguishing NEDE from the earlier Early Dark Energy  model (EDE) \cite{Karwal:2016vyq, Poulin:2018cxd, Poulin:2018dzj} is that normally EDE is realized in terms of a single scalar field that transitions from a slow-roll to an oscillating (or fast-roll) phase via a second-order phase transition, whereas the NEDE is based on a first-order phase transition realized by a quantum tunneling process.

Both single-field EDE and NEDE share two defining properties, which are crucial for their phenomenological success. First, there is an additional energy component, not present in $\Lambda$CDM, which comes to contribute an important fraction to the energy budget at some time $t=t'$ close to matter-radiation equality. Second, that component starts to decay at least as fast as radiation after the time $t'$.

 In the NEDE scheme, as it is discussed in Ref.~\cite{Poulin:2018dzj}, is important to prevent the phase transition from happening too early, in which case the sound horizon and, hence, also $H_0$ would be not affected. Also we need that the phase transition occur on a timescale which is short compared to the Hubble expansion. This avoids the premature nucleation of bubbles of true vacuum that would grow too large before they collide with their smaller cousins. This would lead to large scale anisotropies which would have imprinted themselves in the CMB.

To satisfy these two conditions, NEDE models must include a triggering mechanism for the nucleation process. For example in Refs.~\cite{Niedermann:2019olb, Niedermann:2020dwg} is consider a two-field scalar model in a dark sector that features a built-in trigger mechanism. 

In our model, the time dependence of the scale factor will serve as the driving force behind our triggering mechanism. This is because the total potential, $V_T(\phi_1, \bar{a})$, experienced by the tunneling scalar field, $\phi_1$, depends on the scale factor.

\begin{figure}[t]
\centering
\subfigure{\includegraphics[scale=1.0]{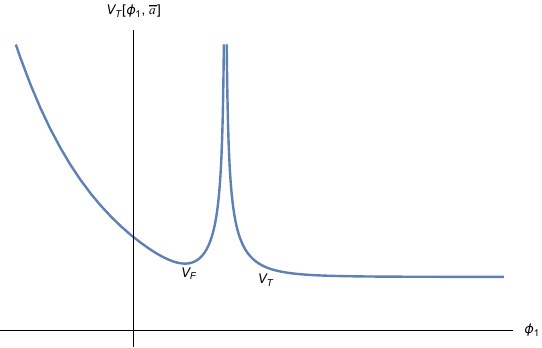}}
\caption{Schematic representation of the total potential $V_T(\phi_1, \bar{a})$ where we have assume that the value of $\bar{a}$ is fixed and the field $\phi_1$ is varied.}
\label{FVT1}
\end{figure}

In particular, the total potential is given by Eq.~(\ref{VT}) and it is show in Fig.~\ref{FVT1}. We can note that the potential presents  a divergence at $\phi_1=\phi_1^0$, where $F(\phi_1^0) \rightarrow \infty$ and $ U_{eff}(\phi_1^0) = 0$. The divergent point is given by, see Eq.~(\ref{F1}) 

\begin{equation}\label{Phi0}
\phi_1^0 = \frac{\log \left(-\frac{f_1}{g_1}\right)}{\sqrt{\alpha_1^2+\alpha_2^2}}\,.
\end{equation}

Thus, the divergent barrier at $\phi_1=\phi_1^0$ separates the false vacuum from the true vacuum. Initially, the field $\phi_1$  is in the false vacuum and then, through a tunneling effect, transitions to the true vacuum. 

We can note from Eq.~(\ref{VT}), that  $V_T(\phi_1, \bar{a})$  depends on cosmic time through its dependence on the scale factor. Consequently, we obtain a decay rate that varies with cosmic time. This provides the necessary mechanism to create a model in the style of NEDE, similar to those studied, for example, in Refs.~\cite{Niedermann:2019olb, Niedermann:2020dwg}, but where the triggering mechanism is driven by the scale factor rather than an additional sub-dominant trigger field.

The tunneling rate per unit volume can be expressed as follows, see \cite{Coleman:1977py}

\begin{equation}\label{gamma}
    \Gamma =  \tilde{A}\, e^{-B/\hbar}\, \big[1 + \mathcal{O}(\hbar) \big] ,
\end{equation}
where $B$ is given by
\begin{equation}
    B = \frac{27\, \pi^2\, S^4_1}{2\, \epsilon^3} \,,
\end{equation}
with the parameter  $\epsilon = V_F -V_T$ and the quantity $S_1$  given by

\begin{equation}\label{S1}
    S_1 = \int^{\phi^+}_{\phi^-} \sqrt{2V_T(\phi)}d\phi   \,.
\end{equation}

Here $\phi^+$ and $\phi^-$  correspond to the initial and final values of the scalar field across the potential barrier, representing the field configuration from which tunneling begins and into which it proceeds during the transition.

We are going to work in the thin-wall approximation where is consider that $\epsilon$ is small. Following Linde \cite{Linde:1981zj}, we assume that the prefactor $\tilde{A}$ in  Eq.~(\ref{gamma}) corresponds to the  nucleation radius  of the bubble $r$ and this radius is defined as

\begin{equation}
    r = \frac{3S_1}{\epsilon} .
\end{equation}
Thus,  the tunneling rate per unit volume $\Gamma$ becomes
\begin{equation}
    \Gamma\sim   \frac{1}{r^4}\, e^{-B/\hbar} =  \frac{\epsilon^4}{(3 S_1)^4}\, \exp\!\left(-\frac{27\, \pi^2\, S^4_1}{2\, \epsilon^3\hbar}\right).
\end{equation}

We now proceed to calculate the term $S_1$ in our model. For this purpose, we consider that for $\phi_1$ near $\phi^0_1$, and then we can approximated the potential $V_T(\phi_1, \bar{a})$  as follow

\begin{equation}
    V_T(\phi_1, \bar{a}) \approx A \, \frac{1}{\left(\left| f_1 e^{-\sqrt{\alpha_1^2+\alpha_2^2}\,\phi_1} + g_1 \right|\right)^\frac{1}{2}}\,,
\end{equation}
where the quantity $A$ is a function of the scale factor in the Einstein frame, $\bar{a}$,  and it is defined as 

\begin{equation}
   A= A (\bar{a})= \left(\frac{8 \chi_2 \left(f_2 e^{-2\sqrt{\alpha_1^2+\alpha_2^2}\,\phi_1^0} + g_2 \right)}{e^{-\frac{\alpha_2^2}{\sqrt{\alpha_1^2+\alpha_2^2}}\phi_1^0}}\right)^\frac{1}{2}\frac{c}{\bar{a}^3} = A_0\,\frac{c}{\bar{a}(\bar{t})^3}\,,
\end{equation}

in which the constant $A_0$ is given by 
\begin{equation}\label{A0}
A_0 =2 \sqrt{2}\,\sqrt{\chi_2\left(\frac{g_1^2f_2}{f_1^2}+ g_2\right) \left(-\frac{f_1}{g_1}\right)^{\frac{\alpha_2^2}{\alpha_1^2+ \alpha_2^2}}}\,.
\end{equation}
Here we have used that the value of $\phi_1^0$ is given by Eq.~(\ref{Phi0}).  

Then, we arrive at the following approximation for the potential $V_T(\phi_1, \bar{a})$

\begin{equation}\label{VTA}
    V_T(\phi_1, \bar{a}) \approx \frac{A_0}{\sqrt{|f_1|}}\;\frac{1}{\left|e^{-\sqrt{\alpha_1^2+\alpha_2^2}\,\phi_1} + \frac{g_1}{f_1} \right|^{\frac{1}{2}}}\;\frac{c}{\bar{a}(\bar{t})^3}\,.
\end{equation}

In this way,  using Eq.~(\ref{VTA}) in the equation for $S_1$, we obtain for our model that

\begin{equation}
    S_1 = \int^{\phi_1^+}_{\phi_1^-} \sqrt{2V_T(\phi)}d\phi \, \approx \, \frac{[2A]^\frac{1}{2}}{f_1^\frac{1}{4}} \int^{\phi_1^+}_{\phi_1^-}\,\frac{d\phi_1}{\left(\left| e^{-\sqrt{\alpha_1^2+\alpha_2^2}\,\phi_1} - \bar{B} \right|\right)^\frac{1}{4}}=\frac{[2A]^\frac{1}{2}}{f_1^\frac{1}{4}}\,I\,,\label{S1}
\end{equation}
where we have defined $\bar{B} = \left|g_1/f_1\right|$ and the integral $I$ is defined as 

\begin{equation}
    I  = \int^{\phi_1^+}_{\phi_1^-}\,\frac{d\phi_1}{\left(\left| e^{-\sqrt{\alpha_1^2+\alpha_2^2}\,\phi_1} - \bar{B} \right|\right)^\frac{1}{4}}\,.
\end{equation}
Using the change of variables $x$ given by 
\begin{equation}
\begin{array}{ccc}
    x &=& e^{-\sqrt{\alpha_1^2+\alpha_2^2}\,\phi_1} \,,
\end{array}
\end{equation}
we find that the integral can be approximated by 
\begin{equation}
    I \approx \frac{1}{\bar{B}\sqrt{\alpha_1^2+\alpha_2^2}}\,  \int^{x^+}_{x^-}\,
    \frac{dx}{|x - \bar{B}|^{\frac{1}{4}}} \,,\label{I1}
\end{equation}
where $x^{\pm}$ are the solutions of the equation

\begin{equation}\label{Abar}
    \sqrt{\frac{1}{| x-\bar{B}| }}- \bar{A} = 0 \,.
\end{equation}

The constant $\bar{A}$ is related to the values $V_F$ and $V_T$ of the potential $V_T(\phi_1, \bar{a})$. We are working in the thin-wall approximation, then we can consider that $V_F \approx V_T$, and
therefore we can write

\begin{equation}
    V_F \approx V_T= V_0\,\frac{c}{\bar{a}(\bar{t})^3}=\frac{A_0}{\sqrt{|f_1|}}\bar{A}\,\frac{c}{\bar{a}(\bar{t})^3}\,,
\end{equation}
and then we have
\begin{equation}
    \bar{A}= \frac{\sqrt{|f_1|}}{A_0}V_0.
\end{equation}

It follows from Eq.~(\ref{Abar}) that   the solutions become
\begin{equation}
x^- = \frac{\bar{A}^2 \bar{B}-1}{\bar{A}^2}\;,\;\; x^+ = \frac{\bar{A}^2 \bar{B}+1}{\bar{A}^2}\,.
\end{equation}

Thus, the integral $I$ given by Eq.~(\ref{I1}) results

\begin{equation}
    I \approx \frac{1}{\bar{B}\sqrt{\alpha_1^2+\alpha_2^2}}\,\left(\frac{8}{3}\frac{1}{\bar{A}^{\frac{3}{2}}}\right).
\end{equation}

On the other hand, from the definition of $\bar{A}$ we can write
 
\begin{equation}
    V_T(\phi_1^{\pm}, \bar{a}) \approx \frac{A_0}{\sqrt{|f_1|}}\;\bar{A}\;\frac{c}{\bar{a}(\bar{t})^3} = V_0\,\frac{c}{\bar{a}(\bar{t})^3}\,,\label{73}
\end{equation}
and then we find from Eq.~(\ref{S1}) that the integral  $S_1$ takes the form
\begin{equation}
    S_1 = \frac{8}{3}\,\frac{\sqrt{2}\,A_0^2}{|f_1|\bar{B}\,\sqrt{\alpha_1^2+\alpha_2^2}}\,
    \left(\frac{1}{V_0}\right)^{3/2}\sqrt{\frac{c}{\bar{a}(\bar{t})^3}}\,.
\end{equation}

Thus, considering the definition of $A_0$, from  Eq. (\ref{A0}), we can express the quantity $S_1$ as

\begin{equation}
    S_1 = A_1\, \sqrt{\frac{c}{\bar{a}(\bar{t})^3}}\,,
\end{equation}
where the quantity $A_1$ is defined as

\begin{equation}
    A_1 =  \frac{\sqrt{2}\,64\,\bar{B}^{-1-\frac{\alpha_2^2}{\alpha_1^2+\alpha_2^2}}}{3\,f_1\,V_0^{3/2}\sqrt{\alpha_1^2+\alpha_2^2}}\left[\big(f_2\bar{B}^2 + g_2\big)\,\chi_2\right] .
\end{equation}

Therefore,  we find that the tunneling rate per unit volume (in units in which $\hbar=1$) for our model can be written as

\begin{equation}\label{gamma2}
    \Gamma = C_1 \bar{a}^6\, e^{-\frac{B_1}{\bar{a}^6}} \,,
\end{equation}
where we have defined the constants $C_1$ and $B_1$ as
\begin{eqnarray}
C_1 &=& \frac{\epsilon^4}{81\,A_1^4\,c^2},\\
\nonumber \\
B_1 &=& \frac{27\,\pi^2\,A_1^4\,c^2}{2 \epsilon^3}\,,\label{c1}
\end{eqnarray}
respectively.

We quantify the efficiency of the bubble nucleation in terms of the percolation parameter $p = \Gamma/\bar{H}^4$, see Refs.~\cite{Niedermann:2019olb, Niedermann:2020dwg, Turner:1992tz}. Provided $p>1$ at least one bubble can be expected to be nucleated within one Hubble patch
and Hubble time. To make the phase transition an instantaneous event on cosmological timescales
and avoid phenomenological problems with large bubbles, we impose the stronger condition $p \gg1$
during bubble percolation. On the other hand, if $p \ll 1$, the percolation cannot keep up with the
expansion of space, and a typical Hubble patch does not contain any bubble and this is the condition that
we want to realize before the transition.

In this context, we have that the percolation parameter for our model is given by

\begin{eqnarray}
    p &=& \frac{\Gamma}{\bar{H}^4}\sim  \frac{r^{-4}}{\bar{H}^4}\, e^{-B/\hbar} 
\nonumber \\
&=& C_1 \bar{a}^6\, e^{-\frac{B_1}{\bar{a}^6}}\frac{1}{\bar{H}^4} = \exp \left\{-\frac{B_1}{\bar{a}^6} + \log \left(\frac{ C_1 \bar{a}^6}{\bar{H}^4}\right) \right\}.\label{pp}
\end{eqnarray}

In order to find a constraint on the parameter $B_1$, we can consider that the percolation time $t'$ occurs when $p(t')\simeq 1$, then from Eq.~(\ref{pp}), we find that the critical value of $B_1$ becomes 
\begin{equation}
B_1\simeq\bar{a}'^6\,\,\mbox{ProductLog}\left[\frac{\epsilon\pi^2}{6\bar{H}'^4}\right],\label{B1}
\end{equation}
where the scale factor $\bar{a}'=\bar{a}(t=t')$, the Hubble parameter $\bar{H}'=\bar{H}(t=t')$ and the  ProductLog function also
called the Omega function or Lambert W function is defined in Ref.~\cite{Prod}.

On the other hand, following Refs.~\cite{Niedermann:2019olb, Niedermann:2020dwg}, we calculate the duration of the percolation phase and provide an estimate for its inverse duration which is given by
\begin{equation}\label{beta}
    \beta= \frac{\dot{\Gamma}}{\Gamma} \,.
\end{equation}

As was mentioned in Ref.~\cite{Niedermann:2020dwg}, this imposes a limit on the maximum time available for bubbles to grow before they begin to collide. Since we require the phase transition to complete within at least one Hubble time, we impose the condition $\bar{H}\beta^{-1}<1$. 

From Eqs.~(\ref{gamma}) and (\ref{beta}), we obtain

\begin{equation}
    \bar{H}\beta^{-1} = \frac{\bar{a}^6}{6(B_1  + \bar{a}^6)}\,.\label{const1}
\end{equation}

We note that the constraint, $\bar{H}\beta^{-1}<1$, is always satisfied in our model.

On the other hand, assuming that in the percolation time $t=t'$,   we can consider that the quantity  $\bar{H}(t=t')\beta^{-1}=
\bar{H}'\beta^{-1}\simeq 10^{-3},
$ \cite{Niedermann:2020dwg} guarantees that the CMB observations do not resolve the spatial structures formed by the largest bubbles.
Then using Eqs.(\ref{B1}) and (\ref{const1}), we find 
\begin{equation}
\frac{\epsilon\pi^2}{6\bar{H}'^4}\simeq10^{2} e^{10^2}\simeq10^{45},\label{rat}
\end{equation}
where we have used that $e^{10^2}\simeq 10^{43}$. This relation will later allow us to determine the parameter 
$\epsilon$, associated with the difference between the respective vacua, provided we can determine the Hubble parameter at the time of percolation  $t'$.

\section{Phenomenological behavior of our model}\label{pheno}

Simulating bubble percolation, along with the subsequent collision and dissipation phases, is a complex task. Therefore, by following Ref.~\cite{Niedermann:2020dwg}, we base our analysis on various simplifying assumptions that separately address the evolution of the background.

In particular we are going to consider that bubble nucleation occurs almost instantaneously on cosmological timescales. In the previous section, we discussed that this requires the condition $H_*\beta^{-1}\ll1$. This condition also ensures that CMB observations do not resolve the spatial structures formed by the largest bubbles (see Ref.~\cite{Niedermann:2020dwg}).

The condensate formed by colliding vacuum bubbles can be described as a fluid with an effective equation of state parameter $\omega_{eff} = 1$ on large scales, see \cite{Barrow:1981pa}.

Then, motivated by the framework of our model, we assume that the onset of the phase transition, occurring at the redshift $z'$, is directly governed by the evolution of the scale factor $\bar{a}$.
Specifically, we propose that the dynamics of the scale factor act as the triggering mechanism for the transition, determining the moment when the system evolves from the false vacuum state to the true vacuum state. This assumption links the phase transition to the underlying cosmological evolution, providing a natural and time-dependent mechanism for initiating the process.

Consistent with these assumptions, we consider that in our model, before the transition ($z > z'$), the matter content consists of radiation, dust (DM and barionic matter), and a cosmological constant $\Lambda_1$, associated with the field $\phi_1$ in its false vacuum. After the transition, for $z < z'$, the matter content includes radiation, dust (DM and barionic matter), and a cosmological constant $\Lambda_2$, associated with the field $\phi_1$ in its true vacuum, as well as a fluid with an effective equation of state parameter $\omega_{eff} = 1$, representing the condensate formed by the colliding vacuum bubbles, as previously discussed.

These assumptions enable us to perform an initial phenomenological assessment of our model. We aim to relax and examine them more thoroughly in future work.

Our effective model can be described in the following way. Before the transition, the Hubble parameter can be written as
\begin{eqnarray}\label{H1}
    \bar{H}_1^2 &=& \frac{\kappa}{3}\Big[ \rho_r(\bar{a}) + \rho_m(\bar{a}) + \Lambda_1 \Big].
\end{eqnarray}

The energy density associated to the dust is given by
\begin{eqnarray}
    \rho_m(\bar{a}) &=& \Big[F(\phi_1) + b_m\,F^{-1}(\phi_1) \Big] \left( \frac{c}{\bar{a}^3}\right)  = \left( 2\sqrt{b_m}  \right) \frac{c}{\bar{a}^3}\,,\\ \nonumber
    \\ 
    \rho_m(\bar{a}) &=& \left(\rho_{DM} + \rho_b \right) \, \frac{1}{\bar{a}^3} = \rho_m \, \frac{1}{\bar{a}^3}\,,
\end{eqnarray}
where $\rho_{DM}$ and $\rho_b$ are the energy densities of dark matter and barionic matter measures today. 

The radiation energy density is composed of photons and neutrinos and is given by
\begin{eqnarray}
    \rho_r(\bar{a}) &=& (\rho_\gamma + \rho_\nu) \, \frac{1}{\bar{a}^4} = \rho_r \, \frac{1}{\bar{a}^4}\,.
     \label{L1}
\end{eqnarray}

Here $\rho_\gamma$ and $\rho_\nu$ are the energy densities of the photons and neutrinos measures today.

On the other hand, as was mentioned, $\Lambda_1$ corresponds to the cosmological constant before the transition and it is given by the effective potential $U_{eff}$ evaluated in the false vacuum $\phi_1^+$, then we have
\begin{eqnarray}
        \Lambda_1 &=& U_{eff}(\phi_1^+) \simeq \frac{f_1^2}{4\chi_2 \,f_2} \; .
\label{L2}
\end{eqnarray}

After the transition, we have
\begin{eqnarray}\label{H2}
    \bar{H}_2^2 &=& \frac{\kappa}{3}\Big[ \rho_r(\bar{a}) + \rho_m(\bar{a}) + \rho_2(\bar{a}) + \Lambda_2\Big].
\end{eqnarray}

Here $\rho_2(\bar{a})$ is the energy density associated to the condensate formed by the colliding vacuum bubbles, with $\omega_{eff}=1$, discussed above, and $\Lambda_2$ is the cosmological constant after the transition to the true vacuum $\phi_1^-$.
Thus, considering  Ref.~\cite{Barrow:1981pa} we have that the energy density $\rho_2(\bar{a})$ is defined as 
\begin{eqnarray}
    \rho_2(\bar{a}) &=& \rho_2 \,\frac{1}{\bar{a}^6}\,, \\
\mbox{and}\,\,\,\,\,  \,\,\,\,\,\,   \Lambda_2 &=& U_{eff}(\phi_1^-)\simeq \frac{g_1^2}{4\chi_2 \,g_2} \,,\label{L2}
\end{eqnarray}  
where $\rho_2$ is a constant, representing the energy densities of the colliding vacuum bubbles condensate, measures today. Moreover, we can recognize that the constant  $\Lambda_2$ is the value of the current cosmological constant. 

We rewrite equations (\ref{H1}) and (\ref{H2}), using the density parameters and  the redshift $z$. In this form, we have that the Hubble parameters $H_1$ and $H_2$ can be rewritten as 
\begin{eqnarray}
    \bar{H}_1^2(z) &=& \bar{H}_0^2 \left[ \Omega_m (1+z)^3 +\Omega_r (1+z)^4 + \Omega_{\Lambda_1}  \right], \\
    \nonumber \\
    \bar{H}_2^2(z) &=& \bar{H}_0^2 \left[ \Omega_m (1+z)^3 +\Omega_r (1+z)^4 + \Omega_{\Lambda_2} + \Omega_{\rho_2} (1+z)^6 \right].
\end{eqnarray}

The transition occurs when $z=z'$ and we have defined $\Omega_m = \frac{\kappa}{3}\frac{\rho_m}{\bar{H}_0^2} = \Omega_{DM}+\Omega_b$, $\Omega_r = \frac{\kappa}{3}\frac{\rho_r}{\bar{H}_0^2}$, $\Omega_{\Lambda_1} = \frac{\kappa}{3}\frac{\Lambda_1}{\bar{H}_0^2}$, $\Omega_{\rho_2} = \frac{\kappa}{3}\frac{\rho_2}{\bar{H}_0^2}$, $\Omega_{\Lambda_2} = \frac{\kappa}{3}\frac{\Lambda_2}{\bar{H}_0^2}$, and $\bar{H}_0$ is the Hubble constant evaluated at the present time.

We assumed the continuity of the background energy density. Then we have the condition, $\bar{H}_1(z')=\bar{H}_2(z')$, at the redshift of transition.

Since at $z=0$ we have 
\begin{equation}
     \Omega_m +\Omega_r + \Omega_{\Lambda_2} + \Omega_{\rho_2} =1 \,,
\end{equation}
then we obtain 
\begin{equation}
     \Omega_{\Lambda_2} =1 -\Omega_m -\Omega_r - \Omega_{\rho_2} \,.
\end{equation}

On the other hand,  by the continuity of the background energy density at the redshift $z=z'$, we have
\begin{equation}
    \Omega_{\Lambda_1} = \Omega_{\Lambda_2} + \Omega_{\rho_2} (1+z')^6 \,.
\end{equation}

Thus, we find  that the density parameter $\Omega_{\rho_2}$ satisfies the following relation
\begin{equation}
     \Omega_{\rho_2}= \frac{\Omega_m +\Omega_r + \Omega_{\Lambda_1} -1}{(1+z')^6 -1} \,.\label{R2}
\end{equation}

Following the scheme of the NEDE models, see Refs.~\cite{Niedermann:2019olb, Niedermann:2020dwg}, we define the parameter $f_{NEDE}$ as follow
\begin{equation}
\frac{\Omega_{\Lambda_1}}{\Omega_m (1+z')^3 +\Omega_r (1+z')^4 + \Omega_{\Lambda_1}} = f_{NEDE}\,.
\label{O1}
\end{equation}

Since our model does not influence the inflationary era or the physics of baryons and radiation, we do not expect any significant deviations in the parameters associated with these sectors.
Therefore, we adopt the values for the spectral index, scalar power spectrum amplitude, etc. as well as the baryonic and radiation components from Planck collaboration \cite{Planck:2018vyg}.

In this sense, we have that  the density parameter related to radiation $\Omega_r$ becomes
\begin{equation}
    \Omega_r = \left(1 + \frac{7}{8}\left(\frac{4}{11}\right)^{\frac{4}{3}}N_{eff}\right)\Omega_\gamma \,,
\end{equation}
where $\Omega_bh^2= 0.02212$, $N_{eff}=3.046$ and $\Omega_\gamma h^2 =  2.469 \times 10^{-5}$, see \cite{Planck:2018vyg}.

Then, at this level, we have the following free parameters that characterize our model, the redshift of the transitions $z'$, the parameter $f_{NEDE}$, the density parameter $\Omega_{m}$ and the Hubble parameter at present $H_0$.

As a first approach to study the plausibility of our model within the NEDE framework, we are going to consider that $z' = 5000$, see \cite{Niedermann:2020dwg} and we allow the dark matter density $\Omega_{m}$, the Hubble constant $H_0$ and $f_{NEDE}$ to be free parameters determined by best fit to observational data.

In this work we are going to use CMB, BAO and local $H_0$ datasets to constrain the model. It is important to mention that for the CMB we have not used the full dataset but the reduced. The reduced CMB data set has been shown to capture the main information in the CMB and is useful for checking models beyond the $\Lambda$CDM, see \cite{Khosravi:2023rhy, CMBRe}. The reduced CMB dataset includes the angular scale of the sound horizon at the last scattering surface $\theta_*$, the CMB shift parameter $R$, the baryon density and the spectral index. 
As was mentioned, since our
model does not affect the spectral index and the baryonic
physics, we do not expect any modification in these
two parameters and we fix them same as their best values
from Planck.
Also, following Ref.~\cite{Khosravi:2023rhy}, we do not use the CMB shift parameter to constrain our model, but, we will show that our final prediction for it is compatible with its value reported by Planck \cite{Planck:2018vyg}.

Then, for BAO data we consider isotropic BAO measurements from  6dFGS~\cite{Beutler:2011hx}, MGS~\cite{Ross:2014qpa}, eBOSS~\cite{eBOSS:2017cqx} and anisotropic BAO measurements from BOSS DR12 \cite{BOSS:2016wmc} and Lyman $\alpha$ forest samples \cite{BOSS:2017uab}.

In particular the isotropic BAO measurements are $D_V(0.106)/r_d = 2.98 \pm 0.13$~\cite{Beutler:2011hx}, $D_V(0.15)/r_d = 4.47 \pm 0.17$~\cite{Ross:2014qpa} and $D_V(1.52)/r_d = 26.1 \pm 1.1$~~\cite{eBOSS:2017cqx}. 

The anisotropic  BAO measurements are $D_A(0.38)/r_d = 7.42$, $D_H(0.38)/r_d = 24.97$, $D_A(0.51)/r_d = 8.85$, $D_H(0.51)/r_d = 22.31$, $D_A(0.61)/r_d = 9.69$, $D_H(0.61)/r_d = 20.49$~\cite{BOSS:2016wmc} and $D_A(2.4)/r_d = 10.76$, $D_H(2.4)/r_d = 8.94$~ \cite{BOSS:2017uab}. The covariance matrix corresponding to the anisotropic BAO data set is taken the same as Ref.~\cite{Evslin:2017qdn}.

The quantity $D_V$ is a combination of the line-of-sight and transverse distance scales defined in Ref.~\cite{SDSS:2005xqv}, $D_M(z)$  is the comoving angular diameter distance, which is related to the physical angular diameter distance by $D_M(z) = (1 + z)D_A(z)$ and $D_H = c/H(z)$ is the Hubble distance. Besides, we define the quantities $D_V(z)$ and $D_A(z)$ as

\begin{equation}
     D_V(z)= \left( D^2_M(z)\frac{z}{H(z)}\right)^{1/3},
\end{equation}

\begin{equation}
     D_A(z)= \frac{1}{(1+z)}\,\int^z_{0} \frac{dz'}{H(z')}\,.
\end{equation}


The comoving size of the sound horizon at the drag epoch is defined as
\begin{equation}
    r_d= \int^ \infty_{z_d} \frac{c_s\,dz}{H(z)}\,,
\end{equation}
where $c_s = 1/\sqrt{3(1+{\cal R})}$ is the sound speed in the photon-baryon fluid, ${\cal R}=\frac{3\Omega_b}{4\Omega_{\gamma}(1+z)}$ \cite{Eisenstein:1997ik} and $z_d$ is the redshift at the drag epoch.

From the CMB, we are going to use the acoustic angular angle $\theta_* = 1.04090 \pm 0.00031$  and the CMB shift parameter $R= 1.7478 \pm 0.0046$ with values reported by Planck \cite{Planck:2018vyg}.
The acoustic angular angle $\theta_*$ is defined as 
\begin{equation}
    \theta_* = \frac{r_s(z_*)}{D_M(z_*)},
\end{equation}  
where $r_s(z_*)$ is the comoving sound horizon at recombination and $D_M(z_*)$ is the comoving angular diameter
distance evaluated at recombination.
The CMB shift parameter $R$ is defined as 
\begin{equation}
    R = \sqrt{\Omega_m H^2_0}D_M(z^*) .
\end{equation}

For direct local measurement of $H_0$ we consider $H_0 = 73.30 \pm 1.04\,\mathrm{km\,s^{-1}\,Mpc^{-1}}$ from \cite{Riess:2021jrx}.

In order to analyze the data and to choose the best fit parameters of our model, we use the Bayesian methods applied to cosmology, see the review Ref.~\cite{Bayesinthesky}. 
In this context, for a set of data $D$ and a model with parameters $\Theta$, parameters estimation can be performed by maximizing the likelihood function 
$\mathcal{L}(D\,|\,\Theta)$, assuming flat priors.
The $68\%$ confidence region ($1\sigma$) corresponds to the set of parameters 
for which $\log \mathcal{L}(D\,|\,\Theta) \geq \log \mathcal{L}_{\text{max}} - \frac{1}{2}$, 
assuming the likelihood is approximately Gaussian near its maximum.

In particular, in this work we consider a multivariate Gaussian likelihood of the form
\begin{equation}
\mathcal{L}(\mathcal{D}|\Theta) = \exp\left(-\frac{\chi^2(\mathcal{D}|\Theta)}{2}\right).
\end{equation}

The $\chi^2$ function, for a set of measurements contained in a vector $\mathcal{S}$, is defined as:
\begin{equation}
\chi^2_{\mathcal{S}} = \left[\mathcal{S}^{\text{obs}} - \mathcal{S}^{\text{th}}\right]^T \mathbf{C}^{-1} \left[\mathcal{S}^{\text{obs}} - \mathcal{S}^{\text{th}}\right],
\end{equation}
where $\mathcal{S}^{\text{obs}}$ represents the measured value, $\mathcal{S}^{\text{th}}$ is the theoretical value computed assuming a model
with parameters $\Theta$ and $\mathbf{C}$ corresponds to the covariance matrix of the measurements contained in the vector $\mathcal{S}^{\text{obs}}$.
In our case, the values in $\mathcal{S}^{\text{th}}$ represent the isotropic BAO measurements $D_V(z)/r_d$; the anisotropic BAO measurements $D_A(z)/r_d$ and $D_H(z)/r_d$; the functions $\theta_*$ for CMB data and the direct local measurement of $H_0$.

\begin{figure}[t]
\centering
\subfigure{\includegraphics[scale=1.0]{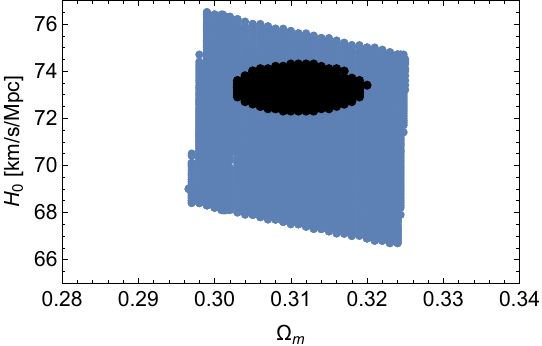}}
\caption{In this plot, the grey region is the $1\sigma$ likelihood for our model which is constrained by only $\theta_*$ + BAO datasets. We can note that it is compatible with the higher local values of $H_0$ which allows us to constrain our model with $\theta_*$ + BAO and $H_0$ datasets. For this case, the black region is the $1\sigma$ likelihood. The black region is plotted in Figure~\ref{H0Om2} in more detail, where the color coding reveals the expected anti-correlation between $H_0$ and $\Omega_m$ at fixed $f_{NEDE}$.}
\label{H0Om1}
\end{figure}

\begin{figure}[t]
\centering
\subfigure{\includegraphics[scale=0.7]{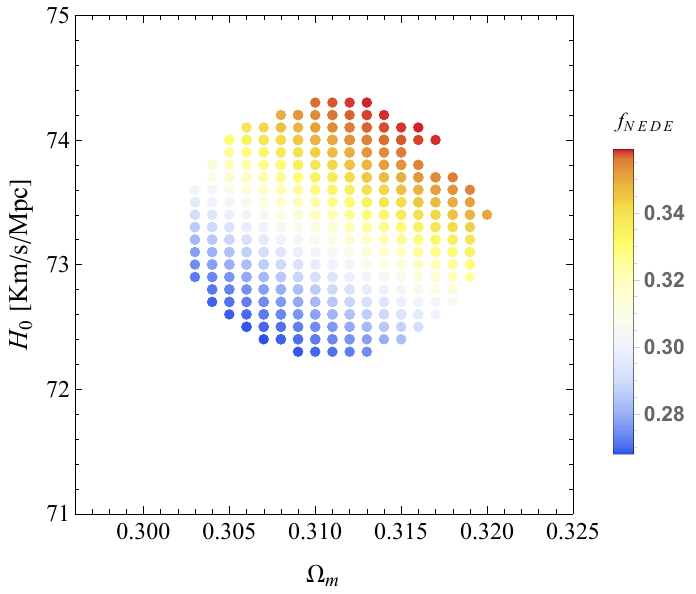}}
\caption{This plot shows the $1\sigma$ contour in the $H_0-\Omega_m$ plane
while the color dots (bar) is representing the $f_{NEDE}$ parameter.
In this plot all the $\theta_*$ + BAO and $H_0$ datasets are used. The
results are totally compatible with higher $H_0$ values while has
no conflict with the $\theta_*$ + BAO. We can notice  that for a fixed
$f_{NEDE}$ value (a fixed color) the expected anti-correlation between
$H_0$ and $\Omega_m$ is seen.}
\label{H0Om2}
\end{figure}

In what follows, we adopt the methodology presented in
\cite{Khosravi:2023rhy}, then we begin by testing our model against the $\theta_*$ + BAO data. If the model proves to be compatible with higher values of $H_0$, we subsequently incorporate the local $H_0$ measurement into the analysis. 
Figure~\ref{H0Om1} presents the results of confronting our model with the $\theta_*$ + BAO dataset, excluding the local $H_0$ measurement. As shown, the resulting posterior $1\sigma$ region is sufficiently broad to accommodate larger values of $H_0$. Accordingly, as the model proves to be compatible with higher values of $H_0$ we constrain our model with the combined dataset $\theta_*$, BAO, and the local $H_0$ measurement. Figure~\ref{H0Om2} displays the $1\sigma$ posterior regions in this case, where the color coding represents the value of $f_{NEDE}$. The results indicate consistency with the higher values of the local $H_0$ measurement, suggesting that our model effectively alleviates the Hubble tension. We note that the standard anti-correlation between $H_0$ and  $\Omega_m$ is clearly visible in Figure~\ref{H0Om2} when $f_{NEDE}$ is held fixed (i.e., for a given color). In Figure~\ref{H0Om1}, this anti-correlation is partially obscured due to the marginalization over $f_{NEDE}$, which introduces an additional degeneracy direction in the $H_0$--$\Omega_m$ plane.

In this analysis, the best-fit values for the free parameters are $H_0 = 73.3 \pm 1\,\mathrm{km\,s^{-1}\,Mpc^{-1}}$, $\Omega_m = 0.311^{+0.009}_{-0.008}$, and $f_{\rm NEDE} = 0.317^{+0.042}_{-0.049}$.
The prediction of the shift parameter for our model, when we use the best fit parameters, is $R = 1.7484$ which is in $1\sigma$ prediction by Planck results.

\subsection{ Model parameters }

In this subsection, we will determine the original parameters of our model, taking into account the observational constraints presented in the previous section. In this sense, we were able to determine the parameters $H_0$, $f_{NEDE}$, and $\Omega_m$ from observations, while keeping the values of $z'$ and $\Omega_r$ fixed.  
Thus, by  combining Eqs.~(\ref{L1}) and (\ref{O1}), we have 
\begin{equation}
\frac{f_1^2}{\chi_2 \,f_2}\simeq \Lambda_1 = U_{eff}(\phi_1^+) =\left[\frac{3H_0^2}{\kappa}\right]\,\left(\frac{f_{NEDE}\,(1+z')^3}{(1-f_{NEDE})}\right)\,[\Omega_m+\Omega_r(1+z')].
\end{equation}
From Eq.(\ref{R2}) we find that the parameter $\rho_2$ associated to the anisotropy energy density is given by
\begin{equation}
\rho_2=\left[\frac{\Omega_m[1+A_f(1+z')^3]+\Omega_r[1+A_f(1+z')^4]-1}{(1+z')^6-1}\right]\,\left(\frac{3H_0^2}{\kappa}\right),
\end{equation}
where the quantity $A_f$ is defined as
$$
A_f=\frac{f_{NEDE}}{(1-f_{NEDE})}.
$$
Similarly, from Eq.~(\ref{L2}) we obtain that
 $$
  \frac{g_1^2}{\chi_2 \,g_2}\simeq \Lambda_2 = U_{eff}(\phi_1^-)=\Bigl[\left(\frac{(1+z')^6}{(1+z')^6-1}\right)+\Omega_m(1+z')^3\left(\frac{A_f+(1+z')^3}{1-(1+z')^6}\right)+
  $$
  \begin{eqnarray}  
  \Omega_r(1+z')^4\left(\frac{A_f+(1+z')^2}{1-(1+z')^6}\right)\Bigr]\,\left(\frac{3H_0^2}{\kappa}\right).
\end{eqnarray}

Now by using the best  observational values for 
$H_0=73.3Km/s Mpc$, 
$\Omega_m=0.311$, 
$f_{NEDE}=0.317$ 
at $z'=5000$ and considering $\Omega_r=0.417698/H_0^2$, we obtain the following values
$$
\frac{f_1^2}{\chi_2 \,f_2}\simeq
7.945\times 10^{-113}M_p^4,\,\,\,\,\,\,\rho_2\simeq
5.078\times10^{-135}M_p^4,\,\,\,\,\,\mbox{and}\,\,\,\,\,\,\,\frac{g_1^2}{\chi_2 \,g_2}\simeq 
1.347\times 10^{-123}M_p^4.
$$

In order to constraint other parameters  using these observational parameter, we consider 
 Eqs.~(\ref{B1}) and (\ref{rat}),  finding that the value of the dimensionless parameter $B_1$ becomes
\begin{equation}
B_1\simeq 6.33\times 10^{-21},\label{B11}
\end{equation}
where we have considered $a'=1/(1+z')$ with the redshift at the percolation time  $z'=5000$. Note that the value of the parameter $B_1$ satisfies the percolation condition $p(t')\simeq $1.

Now by using Eq.~(\ref{rat}), we find that the difference between the false vacuum and the true vacuum $\epsilon$ becomes
\begin{equation}
\epsilon\simeq \,6\times10^{44}\,\,\bar{H}'^4=
6\times10^{44}\,\bar{H}_0^4 \left[ \Omega_m (1+z')^3 +\Omega_r (1+z')^4 + \Omega_{\Lambda_1}  \right]^2\simeq\,
2.711\times10^{-177}M_p^4.
\end{equation}
Here we have considered the Hubble parameter $\bar{H}'=\bar{H}_1(z=z')$ with the redshift $z'=5000$. Additionally, from Eq.~(\ref{c1}) we obtain that the quantity $C_1$ is given by 
\begin{equation}
C_1=\frac{\pi^2}{6}\,\left(\frac{\epsilon}{B_1}\right)\simeq
7.044\times 10^{-157}M_p^4,\label{CC1}
\end{equation}
 we have used Eq.~(\ref{B11}) for the value of $B_1$.
\begin{figure}[t]
    \centering  \includegraphics[width=0.35\linewidth]{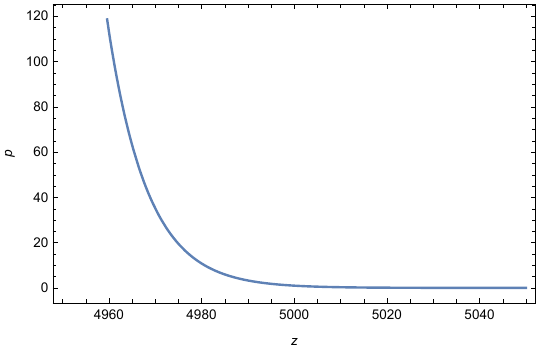}
     \centering  \includegraphics[width=0.35\linewidth]{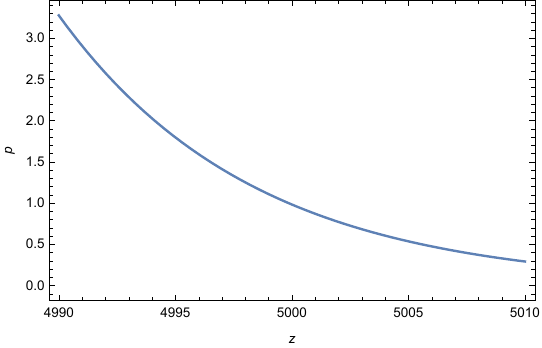}
    \caption{Evolution of the dimensionless percolation parameter $p$ in terms of the redshift $z$. The right panel shows the evolution of the percolation parameter around $z=5000$ where $p(z'=5000)=1$. }
    \label{figp}
\end{figure}

Using these values, we can evaluate the relevant tunneling 
quantities at the transition redshift $z^{\prime}=5000$. 
Since the percolation condition $p(z^{\prime})=1$ implies 
$\Gamma(z^{\prime})=\bar{H}^{\prime 4}$, we obtain a 
tunneling rate $\Gamma(z^{\prime})\simeq 2.8\times 10^{-219}\,
M_p^4$. Furthermore, the ratio between the vacuum energy 
difference $\epsilon$ and the false vacuum energy density 
$\Lambda_1$ yields $\epsilon/\Lambda_1 \sim 10^{-67}$. 
Moreover, since the potential barrier in our model diverges 
at $\phi_1=\phi_1^0$ (see Eq.~(\ref{Phi0})), the condition 
$\epsilon \ll V_{\rm barrier}$ required for the thin-wall 
approximation is automatically satisfied, confirming that 
this approximation is well justified. The bubble nucleation 
radius at the transition satisfies $r/\bar{H}^{\prime -1}
\sim 10^{-10}$, indicating that the nucleated bubbles are 
many orders of magnitude smaller than the Hubble radius, 
which ensures that the phase transition does not introduce 
large-scale anisotropies in the CMB. Additionally, the inverse 
duration of the percolation phase gives 
$\bar{H}^{\prime}\beta^{-1}\simeq 1.7\times 10^{-3}$, 
in agreement with the requirement of Ref.~\cite{Niedermann:2020dwg} and confirming that the transition is effectively instantaneous on cosmological timescales.

In addition, an important requirement for the viability of our model is that the phase transition from EDE to late DE occurs robustly near the desired percolation redshift $z'$ (radiation-matter equality), without fine-tuning of the model parameters. In order to demonstrate this point, we analyze  the sensitivity of the percolation condition to variations in the fundamental tunneling parameter $B_1$. From Eq.~(\ref{B1}) we 
can consider that $B_1\simeq\mathcal{F}(\bar{a}')$ where 
  $\mathcal{F}(\bar{a}')$ is defined as
\begin{equation}
\mathcal{F}(\bar{a}')=\bar{a}'^6\,\,\mbox{ProductLog}\left[\frac{\epsilon\pi^2}{6\bar{H}'^4}\right],\label{B1a}
\end{equation}
where the percolation parameter $p(z')=1$ and  the scale factor $\bar{a}'=1/(1+z')$. The relation $B_1\simeq\mathcal{F}(\bar{a}')$ shows that the redshift $z'$ depends on the parameter $B_1$ logarithmically through the Lambert W or ProductLog function, which implies that the percolation redshift remains remarkably stable under against variations in the quantity $B_1$. In this form, a variation $\Delta B_1$ in the tunneling parameter produces a corresponding variation $\Delta \bar{a}'$ in the percolation scale factor, and hence a variation $\Delta z'$ in the redshift. In order to determine the relation between $\Delta B_1$ and $\Delta \bar{a}'$ (or $\Delta z'$), we differentiate the percolation condition defined by Eq.~(\ref{B1a}), and to first order in $\Delta \bar{a}'$ yields $\Delta B_1\simeq (d\mathcal{F}(\bar{a}')/d\bar{a}')|_{\bar{a}_{(0)}'}\,\Delta \bar{a}'$, where the derivative of $\mathcal{F}(\bar{a}')$ is evaluated in the best-fit percolation scale factor $\bar{a}_{(0)}'=1/(1+z_{(0)}')$.  Adopting  the radiation-dominated approximation, valid around $z'\sim 5000$ one finds approximately that $d\mathcal{F}(\bar{a}')/d\bar{a}'\sim 6\bar{a}'^5\,\mbox{ProductLog}[\epsilon\pi^2/(6\bar{H}'^4)]=6\bar{a}'^5(B_1/\bar{a}'^6)=6B_1/\bar{a}'$, and then the variation $\Delta B_1$ becomes
\begin{equation}
\Delta B_1\simeq \frac{6B_1}{\bar{a}_{(0)}'}\,\Delta\bar{a}'\,\,\,\Rightarrow\,\,\,\,|\Delta z'|\simeq \left(\frac{1+z_{(0)}'}{6}\right)\,\frac{|\Delta B_1|}{B_1},\label{Cont1}
\end{equation}
where we have assumed that the term associated to the variation of the ProductLog function with respect to the scale factor is negligible during this epoch and  we have also used that $\Delta \bar{a}'/\bar{a}_{(0)}'=-\Delta z'/(1+z_{(0)}')$.
In this form, considering $B_1=6.33\times10^{-21}$, the best-fit value of the tunneling parameter obtained from observational constraints at $z'=z_{(0)}'=5000$, and allowing for fractional variation of $\Delta B_1/B_1=0.25$ (the tunneling parameter related to $B_1$ takes a value that is 25$\%$
larger than its best-fit value), we find a variation $\Delta z'=208$, corresponding to a shift of less than 5$\%$ in  the percolation redshift. For a variation $\Delta B_1/B_1=0.10$, corresponding to a 10$\%$
increase relative to its best-fit value, we obtain a  redshift variation $\Delta z'=83$, which  corresponds to a shift of approximately 2$\%$ in the percolation redshift. 
In this sense, these results demonstrate that the triggering of the phase transition near radiation-matter equality is a robust prediction of the model, and is not sensitive to fine-tuning of $B_1$.

The Fig.~\ref{figp} shows the evolution of the percolation parameter versus the redshift $z$ defined by Eq.~(\ref{pp}). The right panel shows the evolution of the percolation parameter versus the redshift in the vicinity $z'=5000$. Here we have utilized the values of the constants $B_1$ and $C_1$ given by Eqs.~(\ref{B11}) and (\ref{CC1}), respectively. From this figure we note that for high redshift in which $z\gg 5000$  the parameter $p\ll 1$. In this sense,  the percolation cannot keep space with the expansion of space since $p\ll1$, and a typical Hubble patch remains devoid of bubbles. On the other hand, for lower redshift $z\ll 5000$, we find  that the percolation parameter $p\gg 1$. Thus,  we ensure that the phase transition occurs as an effectively instantaneous event on cosmological time scales and to avoid phenomenological issues associated with large bubbles \cite{Niedermann:2020dwg}. In this plot, we have considered that the corresponding percolation redshift $z'$ takes place at $z'=5000$ (see Ref.~\cite{Niedermann:2020dwg}) and is implicitly defined by the condition $p(z=z')=1$, which also determines the value of the parameter $B_1$ defined by Eq.~(\ref{B1}).

\begin{figure}[t]
         \subfigure{\includegraphics[scale=0.3]{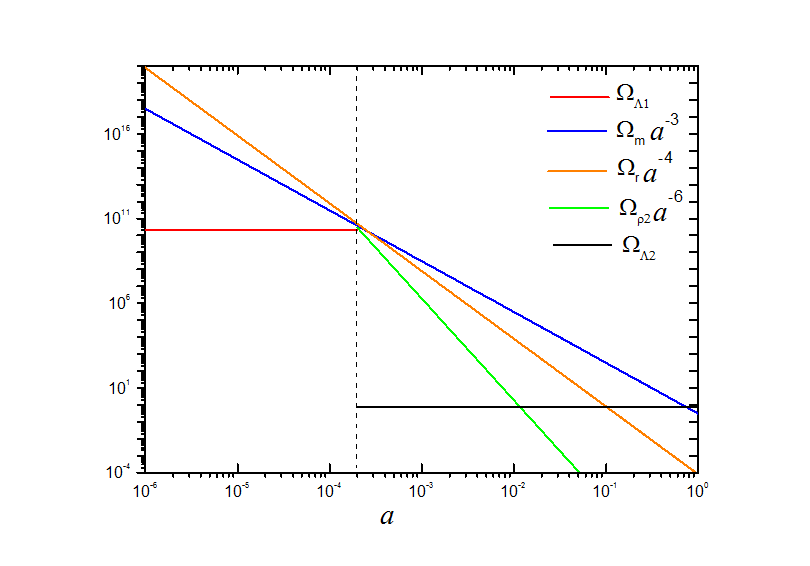}}
    \caption{Evolution of the different fluid components in terms of the scale factor. We have used logarithmic scales on both axes and the dashed vertical line corresponds to $z' = 5000$.}
    \label{fig7}
\end{figure}


Figure \ref{fig7} shows the evolution of the different fluid components as a function of the scale factor. We have also plotted the density parameter associated with the cosmological constant before the transition, $\Omega_{\Lambda_1}$ (red line) and the constant density parameter corresponding to its present-day value, $\Omega_{\Lambda_2}$ (black line). In this analysis, we used the values of the best-fit parameter obtained previously from observational data. 

As mentioned earlier, the functional form of the scalar field potentials is not arbitrary but follows from the requirement of global Weyl-scale invariance of the underlying action. In relation to the presence of several parameters, their main role is to set the relative energy scales associated with the different flat regions of the effective potential $U_{\rm eff}$, see Eq.~\rf{early-vs-late}. These regions can be interpreted as corresponding to the inflationary phase, an early dark energy scenario and the late-time dark energy stage. In particular, the qualitative structure of the potential, characterized by the presence of three quasi-flat regions, emerges within the model and is maintained for a reasonable range of parameters values when  we considered  different large values of the fields $\varphi_1$ and $\varphi_2$. 

Additionally, the parameter values utilized in the cosmological analysis are constrained through a Bayesian fit to observational data from BAO, the local measurement of $H_0$,  and the reduced CMB dataset. In this way, by exploring the region of parameter space allowed by observations, we obtain constraints on the model parameters, ensuring  that the main predictions of the model remain consistent. In this sense, the scenario allows for a transient early dark energy component that reduces the sound horizon and later evolves into the present dark energy regime through the nucleation of bubbles associated with a 
 controlled tunneling process in the effective potential, remain robust throughout the observationally viable region of parameter space. 


\section{Discussion}
\label{discuss}

In the present paper we have constructed a new kind of gravity-matter theory defined in terms of two different non-Riemannian volume-forms (generally covariant integration measure densities) on the space-time manifold. We also introduced two scalar fields in a scale invariant way. The integration of the equations of motion of the degrees of freedom that define the measures provides the constants of integration $M_1$ and $M_2$ which provide us with the spontaneous breaking of scale invariance.

We discover the possibility of three flat regions of the scalar field potential. We have studied this model in what concerns to inflation era by using the highest flat region and its slow roll toward the second flat region. In this paper we identify this second flat region as the Early Dark energy state, for describing the two remaining flat regions we ignore the constants of integration  $M_1$ and $M_2$ and the scalar field potential now depends only on $\phi_1$ allows two remaining different flat regions for possible dark energy sectors. In each of these sectors there are particular values of $\phi_1$ where the matter induces a potential for $\phi_2$ is stabilized. At those points the matter behaves canonically, i.e. the dust does not produce pressure, etc., but in these two different regions the point particle masses are the same. Besides, the scalar field $\phi_2$ remains a massless field in the two flat regions.

The above implies that the two flat regions at the values of $\phi_1$ where the matter behaves canonically  contain the following three elements: a constant DE, a DM component and a massless scalar field, the DE components differ in the two different regions, but concerning the DM,  we have shown that the mass of the DM particles is the same at the minima of the density dependent effective potential, although it can have an up and down jump along the surface of the bubbles that separate the Early DE regions from the late DE regions. Thus, we have calculated the evolution of the Universe in those phases, excluding the transition regions between the two phases using this fact. 

Because of the scale invariant coupling of the scalar fields to dust particles a scalar field potential that depends on the
matter density is generated and a barrier between the Early DE, with a higher energy density and the late DE with a lower energy density  exists, but as the matter gets diluted,  nucleation of bubbles of late DE in the midst of the early DE becomes more probable, until we reach the percolation point, where all the space becomes full of the late DE.
Thus, a key feature of our model is the direct dependence of the tunneling rate on the scale factor, which acts as a natural trigger for the phase transition without requiring additional fields, like in other models of New Early Dark Energy  \cite{Niedermann:2019olb, Niedermann:2020dwg}. The percolation parameter 
$p$ evolves from $p\ll 1$ at high redshifts to $p\gg1$ post-transition, ensuring the process is instantaneous on cosmological timescales and avoids large-scale anisotropies in the CMB.
 
Our model addresses the Hubble tension by modifying the sound horizon prior to recombination  through the introduction of an EDE component that contributes approximately $30\%$ of the energy density around matter-radiation equality. This early injection of energy reduces the sound horizon, allowing for a higher inferred value of $H_0$ consistent with local measurements while preserving agreement with reduced CMB and BAO data. Bayesian analysis of these datasets confirms the viability of our scenario, with the best-fit parameters yielding $H_0 = 73.3 \pm 1\,\mathrm{km\,s^{-1}\,Mpc^{-1}}$, $\Omega_m = 0.311^{+0.009}_{-0.008}$, and $f_{\rm NEDE} = 0.317^{+0.042}_{-0.049}$.

In this initial analysis, we have constrained our model 
using Bayesian methods and observational data from BAO, 
the local measurement of $H_0$, and the reduced CMB 
dataset, following the methodology of Ref.~\cite{Khosravi:2023rhy}. Our 
results indicate that the fraction of Early Dark Energy 
at the time of the phase transition ($z^{\prime}=5000$) 
is approximately 30\% of the total energy density, which 
is higher than the values reported in other NEDE scenarios 
(see Ref.~\cite{Niedermann:2020dwg}). We emphasize that this estimate should 
be regarded as preliminary, since the reduced CMB dataset 
captures the main geometric information of the CMB but 
does not incorporate the full constraining power of the 
complete temperature and polarization power spectra. The 
inclusion of the full CMB likelihood is expected to 
tighten the constraints on $f_{\rm NEDE}$ and likely 
shift the preferred value to smaller fractions, more in 
line with other NEDE analyses in the literature. 
Nevertheless, the primary goal of the present work is to 
establish the viability of the underlying theoretical 
mechanism --- namely, the matter-density-driven tunneling 
from early to late dark energy within a scale-invariant 
framework --- rather than to provide definitive parameter 
constraints. The reduced CMB dataset, combined with BAO 
and local $H_0$ measurements, is well suited for this 
purpose, see Refs.~\cite{Khosravi:2023rhy, CMBRe}. In addition, we 
have employed these best-fit parameters to place 
constraints on the various model-specific quantities, 
such as the tunneling rate, the bubble nucleation radius, 
and the percolation parameter, thereby confirming the 
self-consistency of the tunneling mechanism.

A distinctive feature of our model is 
that the tunneling is triggered by the dilution of matter 
through the scale factor, rather than by an additional 
sub-dominant trigger field, while the underlying 
scale-invariant Lagrangian provides a unified description 
connecting inflation, early dark energy, and late-time 
dark energy within a single framework. An interesting 
consequence of this mechanism is that the transition 
redshift $z^{\prime}$ is not a free parameter but is 
determined by the percolation condition 
$p(z^{\prime})=1$, which depends on the matter content 
through the scale factor. This introduces a correlation 
between $z^{\prime}$ and $\Omega_m$ that is not present 
in NEDE models based on independent trigger fields, and 
which could in principle serve as a distinguishing 
observational signature. Furthermore, the exponential 
dependence of the tunneling rate on $\bar{a}^6$ produces 
a very sharp transition profile (see Figure~\ref{figp}), whose 
specific shape could leave a distinct imprint on the CMB 
compared to other NEDE scenarios with different trigger 
mechanisms. 

In addition, from Eq.~(\ref{Cont1}), we have shown 
the viability of the phase transition from EDE to late-time DE, which occurs robustly near the desired percolation redshift $z'$, without the need for fine-tuning of the model parameter. In this context, the relation between $\Delta z'$ and $\Delta B_1$ allows us to quantify how sensitive the percolation redshift is to variations in the parameter $B_1$, thereby providing a clear measure of the robustness of the tunneling mechanism.

In future work, we plan to constrain the model using the full CMB likelihood, which is expected to further tighten the bounds on $f_{\rm NEDE}$, and to explore in detail the potential observational signatures discussed above.

\begin{acknowledgements}
E.G. want to thank the Pontificia Universidad Cat\'olica de Valpara\'{\i}so, Chile,  for hospitality during this collaboration, and CosmoVerse • COST Action CA21136 Addressing observational tensions in cosmology with systematics and fundamental physics for support for work on this project at BASIC in Ocean Heights, Stella Maris, Long Island,  and to CA23130 - Bridging high and low energies in search of quantum gravity (BridgeQG) for additional support.

P. L. was partially supported by Direcci\'on de Investigaci\'on y Creaci\'on Art\'{\i}stica de la Universidad del B\'{\i}o-B\'{\i}o through Grants RE2320212 and GI2310339.
\end{acknowledgements}


\end{document}